\documentclass[11pt]{article}
\usepackage{graphicx}
\usepackage{multicol,multirow}
\usepackage{amsmath,amssymb,amsfonts}
\usepackage{mathrsfs}
\usepackage{amsthm}
\usepackage{rotating}
\usepackage{authblk}
\usepackage{appendix}
\usepackage{ifpdf}
\usepackage[T1]{fontenc}
\usepackage{newtxtext}
\usepackage{newtxmath}
\usepackage{textcomp}
\usepackage{xcolor}
\usepackage{color}
\usepackage{lipsum}
\usepackage[colorlinks,allcolors=blue]{hyperref}

\usepackage[normalem]{ulem}
\usepackage[margin=1.0in]{geometry}
\normalsize \baselineskip 16pt

\theoremstyle{definition}

\numberwithin{equation}{section}

\newcommand {\beq} {\begin{equation}}
\newcommand {\eeq} {\end{equation}}
\newcommand {\eps} {\varepsilon}

\newcommand {\w} {\overline{w}}
\newcommand {\wh} {\hat{w}}
\newcommand {\vb} {\overline{v}}

\newcommand {\R}{\mathbf{r}}

\newcommand {\A}{\mathbb{A}}

\begin{document}

\title{Whitham modulation equations for the regularized Boussinesq equation with cubic nonlinearity}
\author[1]{Mark A.~Hoefer}
\author[2]{Anna Vainchtein}
\affil[1]{Department of Applied Mathematics, University of Colorado,
  Boulder, Colorado 80309}
\affil[2]{Department of Mathematics, University of Pittsburgh,
  Pennsylvania 15260}
\maketitle

\begin{abstract}
  A regularized Boussinesq equation is studied as a dispersive,
  long-wave (quasicontinuum) approximation of the Fermi-Pasta-Ulam
  lattice with a general cubic interaction force. Explicit periodic
  traveling wave solutions in terms of Jacobi elliptic functions are
  classified, and their solitary-wave, kink, and trigonometric limits
  are obtained.  The Whitham modulation equations describing slow
  modulations of periodic traveling wave solutions are derived using
  an averaged variational principle.  The convexity (strict
  hyperbolicity, genuine nonlinearity) of the resulting
  hydrodynamic-type equations is examined numerically in general and
  analytically in the solitary-wave and harmonic limits.  In
  particular, the loss of hyperbolicity and the formation of
  complex conjugate characteristic velocities is shown to lead to
  modulational instability of periodic traveling waves. The onset of
  modulational instability is verified by numerical computations of
  linearized spectra for periodic traveling waves and initial value
  problems that also reveal additional short-wavelength instabilities.
\end{abstract}

\noindent Competing interests: The authors declare none.\\

\section{Introduction}
\label{sec:intro}
Nonlinear waves in non-integrable Hamiltonian lattices, such as the
celebrated Fermi-Pasta-Ulam (FPU) problem \cite{FPU55} and its various
extensions and physical applications, have attracted a lot of
attention
\cite{Pankov05,Sen08,FlachGorbach08,Kevrekidis11,Chong18,Vainchtein22}. The
dynamics of these waves is very rich due to the combination of higher
order dispersion intrinsic to the lattice setting and higher order
nonlinearity that is often present in such models. In the absence of
integrability, this combination makes the tools of Whitham modulation
theory and more recent advances in dispersive hydrodynamics
\cite{whitham_linear_1999,Kamchatnov00,El16} particularly promising
for providing insight into the collective, multiscale dynamics of
interacting lattice waves and describing complex waveforms such as
dispersive shock waves and traveling waves connecting periodic
orbits. Some work in this direction has begun
\cite{Filip99,Dreyer05,Dreyer06,Dreyer08,Sprenger24,Yang24} but much
remains to be investigated.

A fundamental concept of primary importance for lattice dispersive
hydrodynamics is the hyperbolicity of the modulation equations for the
FPU problem.  Hyperbolicity governs the dynamical stability and
structure of multiscale wave dynamics \cite{whitham_linear_1999}.
In the non-integrable FPU problem, hyperbolicity has only been assessed in
some special limits, such as the harmonic chain, the hard-sphere model
and small-amplitude periodic traveling waves \cite{Dreyer05}.
Another, related concept for a hyperbolic system is that of a
genuinely nonlinear characteristic field \cite{lax_hyperbolic_1973},
which is a necessary condition for the construction of simple waves
(integral curves) that form the foundation for solutions to Riemann
problems and dispersive shock waves \cite{El16}.  We say that a
hydrodynamic-type system of Whitham modulation equations is
\textit{convex} if it is strictly hyperbolic and genuinely nonlinear
\cite{levermore_hyperbolic_1988}.  A modulation system may be convex
for a restricted subset of the state variables.

The present work focuses on studying the
modulation system and its convexity for the modified regularized
Boussinesq equation, a non-integrable dispersive partial differential
equation that approximates FPU dynamics \cite{Rosenau86}.  The
regularized Boussinesq equation has been shown to capture some FPU
features, including traveling kink fronts, solitary waves and
dispersive shocks \cite{Vainchtein24,Yang24}. We consider a general
cubic nonlinearity that enables construction of explicit periodic
traveling waves in terms of elliptic functions and systematically
investigate all of the different cases and special limits.  While it
is typically assumed that all of the roots of the polynomial
determining periodic traveling waves are real, our analysis includes
the case when two of the roots become complex, which is relevant for
the nonlinear interaction force with positive cubic-term coefficient. The
transition from real to complex roots involves a bifurcation where the
solution is a trigonometric wave. We then proceed to derive the
Whitham modulation equations using Whitham's averaged variational
principle \cite{whitham_linear_1999,Kamchatnov00} and discuss their
structure for three distinct cases: i) quadratic nonlinearity $w+w^2$
and cubic nonlinearities ii) $w+w^3$, iii) $w-w^3$.  The cubic cases
correspond to a nonconvex hyperbolic flux.  We also discuss the
reduction of these equations in the harmonic and solitary-wave
limits. In the case of $w-w^3$, we obtain the wave-action
conservation law for the kink limit, which yields a non-classical,
undercompressive shock \cite{Herrmann10,el_dispersive_2017}, or superkink
\cite{gorbushin_transition_2022}.

The main part of the paper focuses on determining the regions of
convexity for the Whitham system where it is strictly hyperbolic and
genuinely nonlinear. In the harmonic and solitary-wave limits, the
convexity is examined analytically, and the full Whitham system is
studied numerically. In each case, we identify parameterizations that
allow us to present the results in a systematic way in terms of
physically relevant parameters. In the case of quadratic nonlinearity,
a symmetry of the governing equations enables a complete convexity
characterization of the modulation system in terms of a
two-dimensional parameter space. Meanwhile, in the cubic cases a
three-dimensional parameterization is needed, and we convey the
results in a two-parameter plane for different values of the wave
amplitude. The regions where hyperbolicity fails are indicated by
contour plots of the imaginary part of the corresponding eigenvalue,
while the loss of genuine nonlinearity is indicated by curves in the
two-dimensional parameter plane.

Our results show that in the case of quadratic nonlinearity, the
Whitham system is convex, i.e., strictly hyperbolic and genuinely
nonlinear, for sufficiently small amplitude and sufficiently large
mean strain.  In the real-root case for nonlinearity $w+w^3$, our
computations suggest strict hyperbolicity of the Whitham equations in
the entire parameter range. However, the system loses genuine
nonlinearity along two curves that bifurcate from the solitary-wave
limit. The situation is different in the complex-root case, where we
observe loss of hyperbolicity in an amplitude-dependent part of the
parameter domain, as well as loss of genuine nonlinearity in up to
three characteristic fields. As the amplitude is increased, the
convexity domain becomes larger. In the case of cubic nonlinearity
$w-w^3$, the convexity figures become even more complex. For a fixed
amplitude, there are two domains where hyperbolicity is lost,
including a parameter range in the solitary-wave limit, in addition to
several curves marking the loss of genuine nonlinearity. The convexity
domain is larger for smaller amplitudes in this case.

The loss of hyperbolicity is known to correspond to modulational
instability \cite{whitham_linear_1999}, and rigorously so in some
specific cases
\cite{bronski_modulational_2010,benzoni-gavage_slow_2014,johnson_modulational_2020,clarke_rigorous_2022}.
We verify the onset of modulational instability by numerically
computing the spectrum of the linearized operator using the
Floquet-Fourier-Hill method \cite{DeconinckKutz06} and show that the
slopes of the associated cross structure in the spectrum near the
origin are well approximated in terms of the real and imaginary parts
of the Whitham characteristic velocities. In addition to confirming
the emergence of modulational instability, our spectral calculations
reveal short-wavelength instabilities in both modulationally stable
and unstable cases, including superharmonic instabilities studied in
detail in \cite{Bronski23} for the case of quadratic nonlinearity. We
also conduct numerical simulations initiated by modulationally
unstable periodic traveling waves perturbed by the corresponding
unstable eigenmode. The results confirm the development of
modulational instability along with short-wavelength instabilities
that eventually lead to the blowup of the solution in some cases.

The rest of the paper is organized as follows. In
Sec.~\ref{sec:prelim} we introduce the quasicontinuum model and
discuss parameter reduction, conservation laws and averaging. Explicit
periodic traveling wave solutions, their harmonic, solitary-wave and
kink limits, existence domain and parameterizations are discussed in
Sec.~\ref{sec:trav-wave-solut}. In Sec.~\ref{sec:whith-modul-equat} we
derive the Whitham modulation equations and investigate their
structure, special limits and convexity. In Sec.~\ref{sec:stab} we
verify the onset of modulational instability and explore other
instability modes by examining the spectra of the linearized operator
and conducting numerical simulations. Concluding remarks are found in
Sec.~\ref{sec:conclusions}.

\section{Preliminaries}
\label{sec:prelim}

\subsection{Quasicontinuum model}
\label{sec:setup}
We consider the dispersive quasicontinuum model (modified regularized Boussinesq equation)
\begin{equation}
u_{tt}-\dfrac{1}{12}u_{xxtt}=(f(u_x))_x,
\label{eq:Bous}
\end{equation}
which approximates the FPU problem
\begin{equation}
\ddot{u}_n=f(u_{n+1}-u_n)-f(u_n-u_{n-1})
\label{eq:FPU}
\end{equation} for a chain of masses with displacements $u_n(t)$ and interaction
force $f(w)$. This can be shown \cite{Rosenau86} by rewriting
\eqref{eq:FPU} in terms of strain variables $w_n=u_n-u_{n-1}$, using a
Pad\'e approximation of the discrete operator and integrating the resulting PDE to obtain the displacement formulation \eqref{eq:Bous}.  In what follows, we consider the cubic interaction force
\begin{equation}
f(w)=w+\alpha w^2+\beta w^3,
\label{eq:cubic}
\end{equation}
which corresponds to the $\alpha$-$\beta$-FPU problem \eqref{eq:FPU}.

Observe that if $\beta > \alpha^2/3$ or $\beta < 0$, the change of variables
\[
  u(x,t) = \left ( 1-\frac{\alpha^2}{3\beta} \right )^{1/2} y(x,\tau) -
  \frac{\alpha}{3\beta} x, \quad \tau =
  \left ( 1-\frac{\alpha^2}{3\beta} \right )^{1/2}t,
\]
yields the reduced equation
\[
  y_{\tau\tau} - y_{xx} - \beta\left (y_x^3 \right )_x -
  \frac{1}{12} y_{xx\tau\tau} = 0
\]
that involves no quadratic term in the interaction force.  Because of
this, we can, in the aforementioned regimes, set $\alpha = 0$.  The
additional rescaling $\sqrt{|\beta|} y \to y$ allows one to take
$\beta = \pm 1$ without loss of generality. Similarly, if $\beta=0$
but $\alpha$ is nonzero (quadratic interaction), the rescaling
$\alpha u \to u$ means that we can set $\alpha = 1$.  This change of
variables is reminiscent of the transformation of solutions of the
Gardner equation to solutions of the modified Korteweg-de Vries or
Korteweg-de Vries equations \cite{kamchatnov_undular_2012}.  In what
follows, we will proceed with $\alpha,\beta \in \mathbb{R}$ and either
$\beta > \alpha^2/3$, $\beta < 0$, or $\beta = 0$ and then restrict to
$\alpha = 0$, $\beta = \pm 1$ and $\alpha= 1$, $\beta=0$ when
convenient.

Linearizing \eqref{eq:Bous} about the uniformly deformed
state by setting $u(x,t) = x \w + \epsilon e^{i(kx-\omega_0t)}$ and taking the limit
$\epsilon \to 0$, we obtain the linear dispersion relation
\begin{equation}
  \label{eq:disp}
  \omega_0^2(k,\w) = \frac{k^2 c_s^2(\w)}{1+k^2/12} < 12 c_s^2(\w),
\end{equation}
where 
\begin{equation}
c_s(\w)=(f'(\w))^{1/2}=(1+2\alpha \w+3\beta \w^2)^{1/2}
\label{eq:sonic}
\end{equation}
is the sound (long-wave) speed.

\subsection{Conservation laws}
Equation \eqref{eq:Bous} corresponds to the Lagrangian density
\begin{equation}
\mathbb{L}=\dfrac{1}{2}v^2-\phi(w)+\dfrac{1}{24}w_t^2,
\label{eq:Lagrangian}
\end{equation}
where $w=u_x$ is the strain, $\phi(w)$ is
  the elastic energy density determining the force $f(w) = \phi'(w)$,
and $v=u_t$ is the particle velocity.  The regularized Boussinesq
equation \eqref{eq:Bous} can be written as the first-order system in
conservative form
\begin{equation}
w_t - v_x = 0, \quad v_t-\left (f(w) + \dfrac{1}{12}w_{tt} \right )_x =0.
\label{eq:system}
\end{equation}
Translational invariance of $\mathbb{L}$ implies two additional
conservation laws that are equivalent for smooth solutions
\cite{Gavrilyuk20}. The first one is the energy balance,
\begin{equation}
\dfrac{\partial}{\partial
  t}\left(\phi(w)+\dfrac{v^2}{2}+\dfrac{1}{24}w_t^2\right)-\dfrac{\partial}{\partial
  x}\left[\left(f(w)+\dfrac{1}{12}w_{tt}\right)v\right]=0,
\label{eq:energy_balance}
\end{equation}
where $\phi(w)+\frac{v^2}{2}+\frac{1}{24}w_t^2$ is the energy density, and $-(f(w)+\frac{1}{12}w_{tt})v$ is the energy flux.
The second (Bernoulli-Eshelby) conservation law is
\begin{equation}
\dfrac{\partial}{\partial t}\left(wv+\dfrac{1}{12}w_x w_t\right)-\dfrac{\partial}{\partial x}\left(w f(w)+\dfrac{1}{12}w w_{tt}+\dfrac{1}{2}v^2-\phi(w)+\dfrac{1}{24}w_t^2\right)=0,
\label{eq:BE}
\end{equation}
where $w v+\frac{1}{12}w_x w_t$ is the momentum density, and
$\phi(w)-w f(w)-\frac{1}{24}w_t^2-\frac{1}{2}v^2-\frac{1}{12}w w_{tt}$
is the momentum flux.

\subsection{Averaged Conservation Laws}
\label{sec:aver-cons-laws}

As shown in Sec.~\ref{sec:trav-wave-solut}, under certain conditions, there exists a four-parameter family of periodic traveling wave
solutions
\begin{equation}
  \label{eq:PTW}
  w = w(\theta;\mathbf{r}), \quad v = v(\theta;\mathbf{r}),
\end{equation}
of \eqref{eq:system}
with modulated phase
\[
  \theta = S(X,T)/\epsilon, \quad \theta_x = S_X = k(X,T), \quad
  \theta_t = S_T = -\omega(X,T),
\]
and parameters $\mathbf{r} = (r_1,r_2,r_3,r_4)$ that are allowed to
vary on slow spatio-temporal scales $X = \epsilon x$,
$T = \epsilon t$, $0 < \epsilon \ll 1$:
$\mathbf{r} = \mathbf{r}(X,T)$.  The dispersion relation yields
$k(X,T) = k(\mathbf{r}(X,T))$,
$\omega(X,T) = \omega(\mathbf{r}(X,T))$.  We can formally obtain the
Whitham modulation equations by averaging three conservation laws
\eqref{eq:system}, \eqref{eq:energy_balance} over a $2\pi$ period
according to
\begin{subequations}
  \label{eq:whitham_avg_cons_laws}
  \begin{align}
    \label{eq:WCL1}
    \w_T - \overline{v}_X &= 0, \\
    \label{eq:WCL2}
    \overline{v}_T - \overline{f(w)}_X &= 0, \\
    \label{eq:WCL3}
    \left ( \overline{\phi(w)} + \frac{1}{2}\overline{v^2} +
    \frac{\omega^2}{24} \overline{w_\theta^2} \right )_T - \left (
    \overline{v f(w)} - \frac{\omega^2}{12}\overline{w_\theta v_\theta}
    \right )_X &= 0 ,
  \end{align}
  accompanied by the conservation of waves
  \begin{equation}
    \label{eq:WCL4}
    S_{XT} = S_{TX} \quad \iff \quad k_T + \omega_X = 0 .
  \end{equation}
  An additional conservation law, derivable from the previous set, is
  the average of momentum \eqref{eq:BE}
  \begin{equation*}
    \left(\overline{wv}-\dfrac{1}{12}k\omega \overline{w_\theta^2}
    \right)_T-\left(\overline{w f(w)}+ \dfrac{1}{2}
      \overline{v^2}- \overline{\phi(w)} - \dfrac{1}{24}
      \omega^2\overline{w_\theta^2} \right)_X =0.
  \end{equation*}
\end{subequations}
Here the averaging of a $2\pi$-periodic function $F(\theta)$ is
defined according to
\begin{equation*}
  \overline{F} \equiv \frac{1}{2\pi} \int_0^{2\pi}
  F(\theta)\,\mathrm{d}\theta .
\end{equation*}

\section{Traveling Wave Solutions}
\label{sec:trav-wave-solut}

We seek traveling wave solutions of eq.~\eqref{eq:Bous} in the form
\begin{equation}
u(x,t)=U(\xi), \quad \xi=x-ct,
\label{eq:TWansatz}
\end{equation}
where $c$ is the velocity, and $\xi$ is the traveling wave coordinate.
Substituting \eqref{eq:TWansatz} into \eqref{eq:Bous} and integrating, we obtain
\begin{equation}
c^2\left(w-\dfrac{1}{12}w''\right)=f(w)-\dfrac{A}{2},
\label{eq:ODE}
\end{equation}
where $w=u_x=U'(\xi)$ is the strain, and $A$ is a constant of integration.
Multiplying \eqref{eq:ODE} by $w'(\xi)$ and integrating, we obtain
\begin{equation}
(w')^2=\dfrac{12}{c^2}(-2\phi(w)+c^2w^2+Aw+B),
\label{eq:orbit}
\end{equation}
where $B$ is another integration constant. Equation
\eqref{eq:cubic} corresponds to
\[
\phi(w)=\dfrac{1}{2}w^2+\dfrac{\alpha}{3}w^3+\dfrac{\beta}{4}w^4,
\]
which together with \eqref{eq:orbit} implies that
\begin{equation}
  \begin{split}
    (w')^2&= \dfrac{12}{c^2} \left( -\dfrac{\beta}{2} w^4-
            \dfrac{2\alpha}{3} w^3+ (c^2-1) w^2+ Aw+ B \right)
  \end{split}
  \label{eq:orbit_cubic_gen}
\end{equation}
must hold along periodic orbits. If $\beta \neq 0$, we have \begin{equation}
(w')^2=-\dfrac{6\beta}{c^2}G_4(w), \quad
G_4(w)=(w-w_1)(w-w_2)(w-w_3)(w-w_4),
\label{eq:orbit_cubic}
\end{equation} where $w_i$, $i=1,\dots,4$, are the four roots of the quartic
polynomial $G_4(w)$, numbered in increasing order of their real parts:
$\text{Re} \ w_1 \leq \text{Re} \ w_2 \leq \text{Re} \ w_3 \leq
\text{Re} \ w_4$, followed by increasing order in the imaginary parts.
It will be helpful to note the polynomial coefficients of $G_4(w)$:
\begin{equation}
  \label{eq:G4}
  G_4(w) = w^4 + \frac{4\alpha}{3\beta} w^3 - \frac{2}{\beta}(c^2-1)
  w^2 - \frac{2}{\beta} A w - \frac{2}{\beta} B .
\end{equation}
In what follows, we will assume that at least two adjacent roots are
real, which implies that either all four roots are real, or two are
real and the other two are complex conjugates.  Phase-plane analysis
shows that this assumption is necessary for the existence of
real-valued periodic orbits described by \eqref{eq:orbit_cubic}.

We can solve for a constraint on the roots and express $c$ in terms of
the roots by equating the coefficients of $G_4(w)$ in
\eqref{eq:orbit_cubic} and \eqref{eq:G4} to obtain
\begin{align}
  \label{eq:root_sum}
   -\frac{4\alpha}{3\beta} &= w_1 + w_2 + w_3 + w_4 , \\
  \label{eq:c_vs_roots}
  c^2 &= 1 - \frac{\beta}{2} ( w_1 w_2 + w_1 w_3 + w_1 w_4 + w_2 w_3 +
        w_2 w_4 + w_3 w_4 ).
\end{align}
Consequently, this is a three-parameter family of traveling wave solutions
characterized by, for example, three of the roots.  Both signs of $c$
are possible but the root ordering must be maintained, which when $w_3$ and $w_4$ are real leads to
the constraint
\begin{equation}
  \label{eq:w4}
  w_4 = -\frac{4\alpha}{3\beta} - w_1 - w_2 - w_3 \ge w_3 \quad \iff
  \quad w_3 \le - \frac{2\alpha}{3\beta} - \frac{w_1+w_2}{2} .
\end{equation}
Furthermore, the velocity $c$ in \eqref{eq:c_vs_roots} must be real, so we also require
\begin{equation}
  \label{eq:root_products}
  1 \ge \frac{\beta}{2} ( w_1 w_2 + w_1 w_3 + w_1 w_4 + w_2 w_3 +
  w_2 w_4 + w_3 w_4 ) .
\end{equation}
Equating the two representations \eqref{eq:orbit_cubic} and
\eqref{eq:G4} of $G_4(w)$ also yields expressions for the integration
constants in terms of the roots
\begin{align}
  \label{eq:A}
  A &= \frac{\beta}{2}(w_1w_2w_3 + w_1 w_2 w_4+w_1 w_3 w_4+w_2 w_3
      w_4), \\
  \label{eq:B}
  B &= -\frac{\beta}{2}w_1w_2w_3w_4 .
\end{align}

If $\beta=0$ ($\alpha$-FPU interaction potential), there are only
three roots, all required to be real and numbered in increasing order. In this case,
\eqref{eq:orbit_cubic_gen} reduces to
\begin{equation}
(w')^2 =-\dfrac{8\alpha}{c^2}G_3(w), \quad G_3(w)=(w-w_1)(w-w_2)(w-w_3),
\label{eq:orbit_quad}
\end{equation}
The expanded polynomial of $G_3(w)$ from
eq.~\eqref{eq:orbit_cubic_gen} is
\begin{equation}
  \label{eq:G3}
  G_3(w) = w^3 - \frac{3}{2\alpha}(c^2-1) w^2 - \frac{3}{2\alpha} A w
  - \frac{3}{2\alpha} B .
\end{equation}
Equating the coefficients of $G_3(w)$ in \eqref{eq:orbit_quad} and
\eqref{eq:G3} determines
the velocity-root relation
\begin{equation}
\label{eq:c_vs_roots_quad}
c^2 = 1 + \frac{2\alpha}{3}(w_1+w_2+w_3),
\end{equation}
which requires $\frac{2\alpha}{3}(w_1+w_2+w_3) > -1$ for real $c$.
Thus, we again obtain a three-parameter family of traveling wave
solutions.  The coefficient relations of $G_3(w)$ also determine
\begin{align}
  \label{eq:A_quad}
  A &= -\frac{2\alpha}{3}(w_1w_2+w_1w_3+w_2w_3), \\
  \label{eq:B_quad}
  B &= \frac{2\alpha}{3}w_1w_2w_3 .
\end{align}

Once $\alpha$ and $\beta$ are fixed, the three-parameter family of
traveling waves is parameterized by $(c,A,B)$ or, equivalently, by
$(w_1,w_2,w_3)$.  Below, we will identify an additional, physical
parameterization $(k,a,\w)$ corresponding to the periodic traveling
wave's wavenumber $k$, amplitude $a$, and mean $\w$.

Although we will obtain three-parameter families of periodic traveling
wave solutions for strain $w$, the regularized Boussinesq equation
also requires the determination of particle velocity $v$ in
\eqref{eq:system}, which will introduce another integration constant.
To fix this, we insert the traveling wave ansatz into the first
equation in \eqref{eq:system} and integrate to obtain
\begin{equation}
\label{eq:v_TW}
v(\xi) = -c w(\xi) + V,
\end{equation}
where $V$ is the integration constant.  Averaging the latter
expression over the period $L$ of the traveling wave in terms of
$\xi$, we obtain $V$ as a mean accelerated particle velocity
\[
  V = \vb + c \w = \frac{1}{L}\int_0^L (v(\xi) + c
  w(\xi))\,\mathrm{d}\xi .
\]

The periodic traveling wave solution \eqref{eq:PTW} satisfying
\eqref{eq:orbit_cubic_gen} can be parameterized by the roots
$w_1$, $w_2$, $w_3$ and the mean particle velocity $\vb$.
Collecting these into the vector
\begin{equation}
  \label{eq:w_parameterization}
  \mathbf{W} = (w_1,w_2,w_3,\vb)^T,
\end{equation}
we will later consider alternative mathematically and physically
convenient parameterizations.

We start by considering the general cubic nonlinearity
\eqref{eq:cubic} with $\beta \neq 0$ and the orbit described by
\eqref{eq:orbit_cubic}. First, we observe that in this case
\eqref{eq:ODE} becomes \begin{equation} w''=12\left((c^2-1)w-\alpha w^2-\beta
  w^3+\dfrac{A}{2}\right).
\label{eq:ODE_cubic}
\end{equation}
In what follows, we assume that the right hand side of this
equation has three real roots, corresponding to equilibrium points
$(w,w')=(w_*,0)$ in the phase plane. These equilibria are centers if
$c^2<c_s^2(w_*)$ and saddles if $c^2>c_s^2(w_*)$. Depending on
$\alpha$ and $\beta$, the phase plane of \eqref{eq:ODE_cubic} for given $c$
and $A$ has either two centers separated by a saddle point
(Fig.~\ref{fig:orbits1}(a)) or two saddle points with a center in
between (Fig.~\ref{fig:orbits1}(b)). Hence the number of homoclinic
orbits that enclose families of periodic trajectories is either two or
one, respectively. Below we consider these two cases separately.

\subsection{Case I: $\beta>0$, two centers}
\label{sec:case-i-pos-beta}

\begin{figure}
\centering
\includegraphics[width=\textwidth]{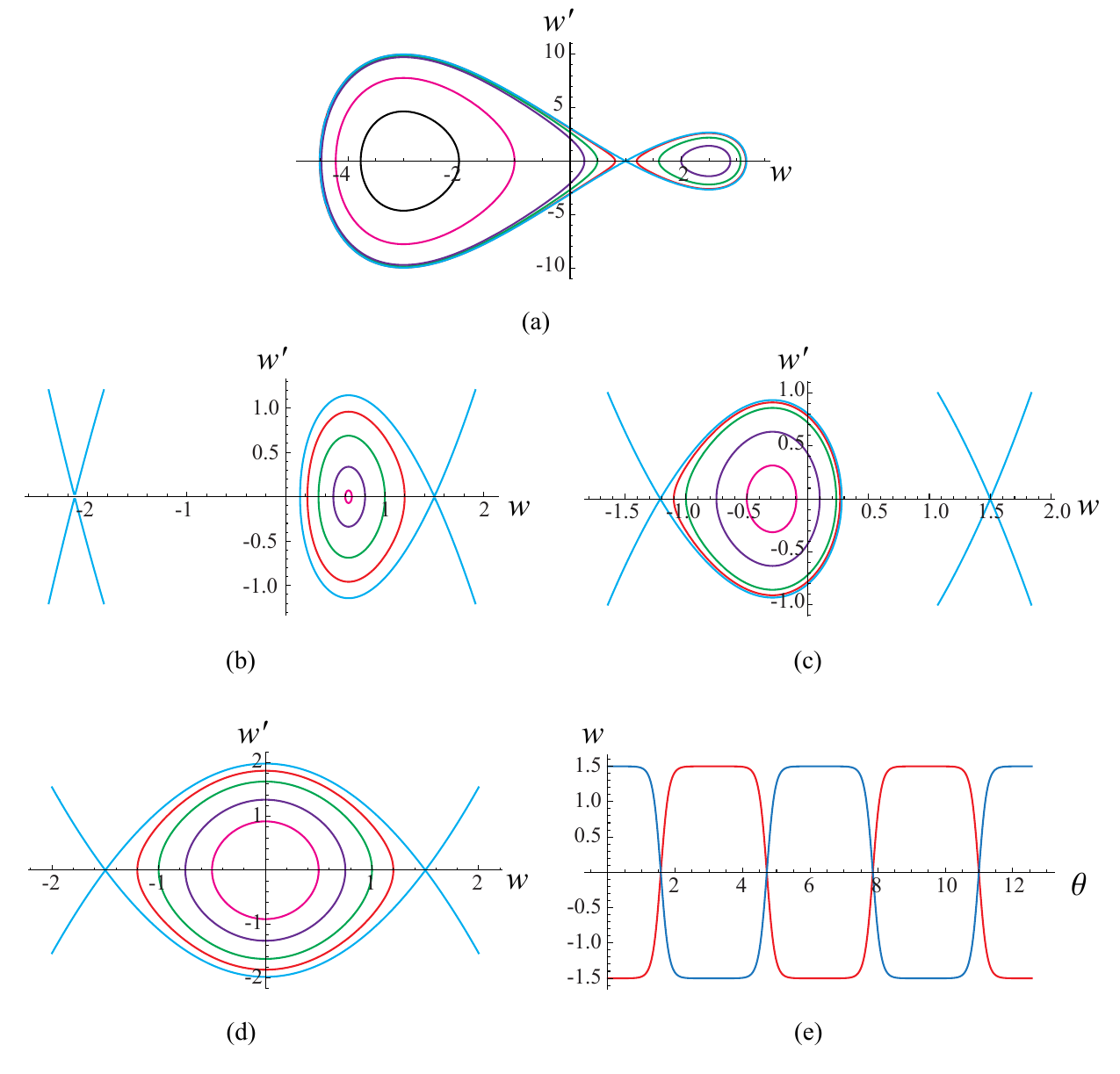}
\caption{(a) Periodic orbits for different values of $B$ at
  $\alpha=0$, $\beta=1$, $c=3$, $A=-127/32$. The orbits on the left of
  the saddle point correspond to solutions \eqref{eq:caseIa} with
  $m=0.442$, $n=0.889$, $a=3.677$ (red), $m=0.216$, $n=0.694$,
  $a=3.07$ (green), $m=0$, $n=0.581$, $a=2.427$ (magenta),
  $n=0.255-0.049i$, $a=1.342$ (black). The orbits on the right
  correspond to the second family of solutions obtained via
  \eqref{eq:sym_gen}, with the same roots $w_i$ as the orbits of the
  same color on the left but amplitude $a=w_4-w_3$ given by $a=2.595$
  (red) and $a=1.779$ (green). (b) Periodic orbits at $\alpha=0$,
  $\beta=-1$, $c=0.85$, $A=11/400$ corresponding to \eqref{eq:caseIIa}
  with $m=0.902$, $n=0.557$, $a=0.638$ (red), $m=0.769$, $n=0.445$,
  $a=0.474$ (green), $m=0.565$, $n=0.297$, $a=0.291$ (magenta),
  $m=0.239$, $n=0.111$, $a=0.099$ (black). (c) Periodic orbits at
  $\alpha=0$, $\beta=-1$, $c=\sqrt{3}/2$, $A=0$ corresponding to
  \eqref{eq:caseIIa} with $m=0.924$, $n=0.726$, $a=0.7$ (red),
  $m=0.796$, $n=0.549$, $a=0.5$ (green), $m=0.4375$, $n=0.25$, $a=0.2$
  (magenta). (d) Solutions \eqref{eq:caseIIa} (red) and its symmetric
  counterpart (blue) in the case (c) near the kink limit $m=n=a=1$. In
  panels (a-c), manifolds associated with the saddle points (including
  homoclinic and heteroclinic orbits) are shown in cyan}
\label{fig:orbits1}
\end{figure}

As illustrated in Fig.~\ref{fig:orbits1}(a), in this case there are
two centers separated by a saddle point and thus two families of
periodic orbits described by \eqref{eq:orbit_cubic}, with $w_1 \leq w \leq w_2$
(below the saddle point) and $w_3 \leq w \leq w_4$ (above it), respectively. The
first family of orbits is enclosed by the homoclinic trajectory
corresponding to a compressive (depression or dark) solitary wave, and
the second by a homoclinic orbit corresponding to a tensile (elevation
or bright) solitary wave.
To obtain the first family, we need to solve \eqref{eq:orbit_cubic}
for $w(\xi)$ in the interval $(w_1,w_2)$, where $w_1$ and $w_2$ must
be real and distinct. Using separation of variables and an appropriate
substitution to evaluate the integral involving $w$ \cite{Byrd13},
we obtain
\begin{equation}
\begin{split}
  w(\theta) &= \frac{w_1+w_4n\, \mathrm{sn}^2\left (
              \frac{K(m)}{\pi} \theta; m \right )}{1 +n\,\mathrm{sn}^2\left (
              \frac{K(m)}{\pi} \theta; m \right )}, \quad
              m = \frac{(w_4-w_3)(w_2-w_1)}{(w_4-w_2)(w_3-w_1)}, \quad n=\dfrac{w_2-w_1}{w_4-w_2},
  \end{split}
\label{eq:caseIa}
\end{equation}
where $\text{sn}$ is the Jacobi elliptic sine function and $K(m)$ is the
complete elliptic integral of the first kind. The solution
\eqref{eq:caseIa} is $2\pi$-periodic, $w(\theta + 2\pi) = w(\theta)$, in
the phase $\theta = k\xi = kx - \omega t$ with wavenumber $k$ and nonlinear frequency $\omega$ given by
\begin{equation}
  \label{eq:k_omega}
  k = \frac{\pi}{K(m)} \sqrt{\frac{3\beta}{2c^2}(w_4-w_2)(w_3-w_1)}, \quad \omega=ck,
\end{equation}
where we recall \eqref{eq:c_vs_roots},
the wavelength $L = 2\pi/k$, and the temporal period $T = 2\pi/\omega$.
The physical parametrization of this solution is $(k,a,\w)$, where the wavenumber $k$ in \eqref{eq:k_omega}, the
amplitude $a = w_2 - w_1$ and the mean
\begin{equation}
\w=\frac{1}{2\pi}\int_0^{2\pi} w(\theta)\,\mathrm{d}\theta = w_4-\dfrac{\Pi(-n,m)}{K(m)}(w_4-w_1)
\label{eq:caseIa_mean}
\end{equation}
are all determined in terms of the roots $w_1$, $w_2$, $w_3$ in view
of \eqref{eq:root_sum}.  The function $\Pi(-n,m)$ is the complete
elliptic integral of the third kind with $n$ given in
\eqref{eq:caseIa}.

Note that the solution \eqref{eq:caseIa} only requires $w_1$ and $w_2$
to be real and also works in the case when $w_3$ and $w_4$ are
non-real complex conjugates. In this case, $m$ and $n$ are complex but
the solution itself and $k$, $a$, $\w$ are real. Example of a periodic
orbit with complex $w_{3,4}$ is the black trajectory in
Fig.~\ref{fig:orbits1}(a). In the case when $w_3$ and $w_4$ are real,
the root ordering $w_1<w_2<w_3<w_4$ implies that
\begin{equation}
n>0, \quad 0<m<1.
\label{eq:nm_bounds_caseI_real}
\end{equation}
If these roots are complex conjugates, $w_{3,4}=\rho \mp i\eta$, $\eta>0$, \eqref{eq:root_sum} implies that $\rho=-(4\alpha/(3\beta)+w_1+w_2)/2$, and the ordering $w_2<\rho$ thus means that $4\alpha + 3\beta (w_1 + 3 w_2)<0$, which together with $w_1<w_2$ yields
\begin{equation}
\begin{split}
&\text{Re}(n)=-\dfrac{6\beta (w_2 - w_1) (4\alpha + 3\beta (w_1 + 3 w_2))}{36 \beta^2 \eta^2 + (4 \alpha + 3\beta (w_1 + 3 w_2))^2}>0, \\
&\text{Im}(n)=-\dfrac{36 \beta^2 \eta (w_2 - w_1)}{36\beta^2 \eta^2 + (4\alpha + 3\beta (w_1 + 3 w_2))^2}<0.
\end{split}
\label{eq:n_bounds_caseI_complex}
\end{equation} As we will show below, in this case we can obtain $m$ in terms of
real and imaginary parts of $n$.  

The bifurcation from real to complex conjugate roots occurs in the
limit $w_3 \to w_4$ where $m \to 0$ so that the solution \eqref{eq:caseIa}
is the trigonometric wave
\begin{equation}
  \label{eq:trig_soln_caseIa}
  w(\theta) = \frac{w_1 + w_4 n \sin^2(\theta/2)}{1 + n
    \sin^2(\theta/2)}, 
\end{equation}
whose wavenumber, amplitude, and mean are
\begin{equation}
  \label{eq:trig_caseIa_k_a_mean}
  k = 2  \sqrt{\frac{3\beta}{2c^2}(w_4-w_2)(w_4-w_1)},  \quad a = w_2
  - w_1, \quad \w = w_4 - \frac{w_4-w_1}{\sqrt{1+n}} .
\end{equation}
Since $2w_4 = -4 \alpha/(3\beta) - w_1 - w_2$ by virtue of
\eqref{eq:root_sum}, the trigonometric wave
\eqref{eq:trig_soln_caseIa} is a two-parameter family of solutions. An
example orbit given by the trigonometric solution
\eqref{eq:trig_soln_caseIa} is the magenta trajectory in
Fig.~\ref{fig:orbits1}.
 
In the limit $w_2 \to w_1$ in eq.~\eqref{eq:caseIa}, $m \to 0$, and
the solution exhibits the following asymptotic
behavior:
\begin{equation*}
  w(\theta) = w_1 - \frac{1}{2}(w_2 -w_1)\cos\theta + \dots ,
  \quad \theta = kx - \omega_0 t,
\end{equation*}
where, upon utilizing \eqref{eq:root_sum}, \eqref{eq:c_vs_roots},
\eqref{eq:k_omega}, and \eqref{eq:caseIa_mean}, we obtain the
linear dispersion relation \eqref{eq:disp} for small-amplitude
(linear) waves propagating on the constant background
$\w = w_1$.

In the limit $w_2 \to w_3$, $m \to 1$, and the solution \eqref{eq:caseIa}
approaches the depression solitary wave
\begin{equation}
  \label{eq:SW1_caseIa}
  w(\xi) = \frac{w_1 + w_4 n\,
    \mathrm{tanh}^2(\xi/\ell)}{1 + n\,
    \mathrm{tanh}^2(\xi/\ell)} , \quad n = \frac{w_3-w_1}{w_4-w_3},
  \quad  \ell = \left (\frac{3\beta}{2c^2}(w_4-w_3)(w_3-w_1) \right
  )^{-1/2},
\end{equation}
with amplitude $a = w_3-w_1$ propagating on the background $\w =
w_3$. An example of the corresponding trajectory is shown by the left
cyan loop in Fig.~\ref{fig:orbits1}(a). These depression solitary
waves form a two-parameter family of solutions with velocity-amplitude
relation
\begin{equation}
  \label{eq:c_SW1_caseIa}
  c^2 = c_s^2(\w) + \frac{1}{2} a^2 \beta-\frac{2}{3} a(\alpha + 3 \beta\w) .
\end{equation}
Strict inequality $w_3<w_4$ and \eqref{eq:root_sum} imply that these waves exist when
\begin{equation}
  \label{eq:SW1_caseIa_existence}
  \w<-\dfrac{\alpha}{3\beta},~ a>0 \quad \text{or} \quad a > 4\left(\w+\dfrac{\alpha}{3\beta}\right)>0,
\end{equation}
which means that $c^2>c_s^2(\w)$ in \eqref{eq:c_SW1_caseIa}, and thus the waves are supersonic. Note that this implies that for $\w>-\alpha/(3\beta)$ the amplitude has a strictly positive \emph{lower bound}.

In the limit $w_2 \to w_3 \to w_4$, we have $\ell \to \infty$, so that
$\tanh^2(\xi/\ell) \sim (\xi/\ell)^2$ in eq.~\eqref{eq:SW1_caseIa}, and the
algebraically decaying depression solitary wave is obtained,
\begin{equation}
  \label{eq:SW2_caseIa}
  w(\xi) = \frac{2 w_1 - 3 \beta w_4 \left ( w_1(w_4+w_1) -
  (w_4-w_1)^2 \xi^2 \right )}{2 - 3 \beta \left ( w_4(w_4+w_1) -
  (w_4-w_1)^2 \xi^2 \right )},
\end{equation}
with amplitude $a = w_4 - w_1$ and background $\w = w_4$. By virtue of \eqref{eq:root_sum}, this is a one-parameter family of algebraic solitary waves subject to the amplitude-background constraint $a = 4(\w + \alpha/(3\beta)) > 0$. These waves propagate with \emph{sound} velocity,
\begin{equation}
  \label{eq:c_SW2_caseIa}
  c^2 = 1 + 3\beta\w(\tfrac{1}{2} a - \w)=c_s^2(\w),
\end{equation}
and thus represent the nontrivial sonic limit of \eqref{eq:SW1_caseIa}
\cite{Vainchtein20}.  The algebraic solitary wave solution
\eqref{eq:SW2_caseIa} can alternatively be obtained by taking the
limit $w_2 \to w_4$ in the trigonometric wave
\eqref{eq:trig_soln_caseIa} because in this limit $k \to 0$ while
$n \to \infty$, so that
$n \sin^2(k\xi) \sim n (k\xi)^2 \to 6\beta(w_4-w_1)^2\xi^2/c^2$.

The second periodic traveling wave family (illustrated by the right
set of orbits in Fig.~\ref{fig:orbits1}(a)) solves
\eqref{eq:orbit_cubic} in $(w_3,w_4)$ with real and distinct $w_3$ and
$w_4$. It is readily obtained from the first family by using the
symmetry of the problem, which amounts to making the following
replacements in \eqref{eq:caseIa}:
\begin{equation}
\begin{split}
&w(\theta) \to -w(\theta)-\dfrac{2\alpha}{3\beta},\\
&w_1 \to -w_4-\dfrac{2\alpha}{3\beta}, \quad w_2 \to -w_3-\dfrac{2\alpha}{3\beta},
\quad w_3 \to -w_2-\dfrac{2\alpha}{3\beta},
\end{split}
\label{eq:sym_gen}
\end{equation}
where the roots are renumbered to keep their nondecreasing order. The
solution has the amplitude $a = w_4 - w_3$ and the same wavenumber
\eqref{eq:k_omega}, wavelength, frequency, and period as the solutions
\eqref{eq:caseIa}. It leads to a two-parameter family of trigonometric
solutions in the limit $w_2 \to w_1$, a two-parameter elevation
solitary wave family in the limit $w_3 \to w_2$, and a one-parameter
family of sonic algebraic elevation waves when $w_3 \to w_2 \to w_1$.

For simplicity, we now consider $(\alpha,\beta)=(0,1)$, recalling that
one can always reduce the case $\beta>\alpha^2/3$ to these parameter
values upon a proper rescaling, as discussed in
Sec.~\ref{sec:setup}. The periodic traveling wave solution
\eqref{eq:caseIa} can be parameterized by $\mathbf{W}$ in
\eqref{eq:w_parameterization}, which involves the first three roots
and the mean particle velocity.  First, we consider the case when all
four roots are real. In this case, \eqref{eq:root_sum} and
\eqref{eq:c_vs_roots} for $\alpha=0$ and $\beta=1$ imply
\[
c^2=1+\dfrac{1}{4}\left((w_1+w_2)^2+(w_2+w_3)^2+(w_1+w_3)^2\right) \geq 1,
\]
so that $c$ is always real (eq.~\eqref{eq:root_products} holds), and
we have $|c|\geq 1$. A second, useful parameterization involves the
elliptic function parameters $n$, $m$, and the wave amplitude $a$:
\begin{align}
  \label{eq:A_parameterization_caseIandII}
  \mathbf{A} = (n,m,a,\vb)^T.
\end{align}


The transformation
$\mathbf{W} = \mathbf{W}(\mathbf{A})$ and its Jacobian determinant are
\begin{equation}
  \label{eq:A_parameterization_transformation_caseIa}
  \begin{split}
    w_1 &= a \Xi(n,m)\gamma_1(n,m), \quad
          w_2 = a \Xi(n,m) \gamma_2(n,m), \quad
          w_3 =  a \Xi(n,m) \gamma_3(n,m), \\
    \Xi &=\frac{1}{4n(m+n)} , \quad
              \det \left (\frac{\partial \mathbf{W}}{\partial \mathbf{A}}
              \right )
              = \frac{a^2(1+n)}{4n^2(m+n)^2} ,
  \end{split}
\end{equation}
where
\begin{equation*}
  \begin{split}
    \gamma_1 &= -3n^2 - 2(1+m)n - m, \quad 
    \gamma_2 = n^2 - 2n(1-m) - m, \\
    \gamma_3 &= n^2 + 2n(1-m) - m .
  \end{split}
\end{equation*}
For $a > 0$ and \eqref{eq:nm_bounds_caseI_real}, the transformation
\eqref{eq:A_parameterization_transformation_caseIa} is invertible. In
these variables, the existence of periodic traveling waves is
restricted by \eqref{eq:nm_bounds_caseI_real}:
\begin{equation}
  \label{eq:A_existence_caseIa}
  \mathbf{A} \in E_a = \left \{  (n,m,a,\vb)^T \in \mathbb{R}^4 \,|\, n>0, ~0< m
      <1,~a>0 \right \}. 
\end{equation}

Another parameterization results by transforming $a$ in $\mathbf{A}$
to the wave frequency
\begin{equation}
  \label{eq:omega_to_a_caseIa}
  \omega = \mathrm{sgn}(c)\frac{\pi a}{K(m)}
  \sqrt{\frac{3(1+n)}{2n(m+n)}} , 
\end{equation}
which is invertible when \eqref{eq:A_existence_caseIa} holds.  Similar
transformations to $k$ and $\w$ can also be obtained.

For the case of complex conjugate roots
$w_{3,4}=-\frac{1}{2}(w_1+w_2) \mp i\eta$, $\eta>0$, it is convenient
to use
\begin{equation}
\mathbf{\tilde{A}}=(n_r,n_i,a,\vb)^T
\label{eq:A_param_caseI_complex}
\end{equation}
as the parametrization, where $n_r=\text{Re}(n)>0$ and
$n_i=\text{Im}(n)<0$ by \eqref{eq:n_bounds_caseI_complex}. This yields
the transformation $\mathbf{W}(\mathbf{\tilde{A}})$
\begin{equation}
\begin{split}
w_1=-\dfrac{a}{4}\left(3 + \dfrac{2 n_r}{n_i^2 + n_r^2}\right), \quad
&w_2=\dfrac{a}{4}\left(1 - \dfrac{2 n_r}{n_i^2 + n_r^2}\right), \quad
w_3=\dfrac{a}{4}\left(1 + \dfrac{2 n_r}{n_i^2 + n_r^2}\right)+i\dfrac{a n_i}{n_i^2 + n_r^2}, \\
&\det \left (\frac{\partial \mathbf{W}}{\partial \mathbf{\tilde{A}}}
              \right )
              = \frac{i a^2}{2(n_i^2 + n_r^2)^2},
\end{split}
\label{eq:transformation_caseI_complex}
\end{equation} that is invertible for $a>0$ with
\begin{equation*}
  m=\dfrac{2 n_i}{n_i + i(1 + n_r)}.
\end{equation*}
Recalling \eqref{eq:n_bounds_caseI_complex} and \eqref{eq:root_products}, we observe that periodic traveling waves exist when
\begin{equation}
  \label{eq:tildeA_existence_caseIa}
  \begin{split}
    &\mathbf{\tilde{A}} \in \tilde{E}_a = \bigg\{  (n_r,n_i,a,\vb)^T
      \in \mathbb{R}^4\,|\, n_r>0, ~n_i<0,~a>0,\\
    &\qquad\qquad\qquad 16 + a^2 \left(3 + \dfrac{4 (n_i^2 (n_r - 2) +
      n_r^2 (1 + n_r))}{(n_i^2 + n_r^2)^2}\right)>0 \bigg\}. 
  \end{split}    
\end{equation}

The transition from real roots to a pair of complex conjugate roots
occurs when $w_3 \to w_4$, so that $m \to 0$.  Passing to this limit in
\eqref{eq:A_parameterization_transformation_caseIa},
\eqref{eq:omega_to_a_caseIa}, and \eqref{eq:trig_caseIa_k_a_mean} for
$(\alpha,\beta) = (0,1)$, we can eliminate $n$ and identify the
crossover dispersion relation
\begin{equation*}
  \omega^2 = k^2 \frac{48 + 9a^2}{48-2k^2}, \quad 0 \le k < 2 \sqrt{6} ,
\end{equation*}
holding for the trigonometric solution \eqref{eq:trig_soln_caseIa}
with wave speed satisfying $c^2 = 1 + \frac{3}{16} a^2 + \frac{1}{24} \omega^2$.

\subsection{Case II, $\beta<0$, single center}
\label{sec:case-ii-neg-beta}

In this case, as shown in Fig.~\ref{fig:orbits1}(b), there are two
saddle points separated by a center, and hence periodic
orbits satisfy \eqref{eq:orbit_cubic} in $(w_2,w_3)$, where
$w_2$ and $w_3$ are real and distinct. One family of periodic traveling waves is
given by \cite{Chong22}
\begin{equation}
  w(\theta)=\dfrac{w_2-w_1n\,\text{sn}^2\left
      (\frac{K(m)}{\pi}\theta;m \right )} 
  {1-n\, \text{sn}^2 \left (\frac{K(m)}{\pi} \theta;m \right )},
  \quad m=\dfrac{(w_4-w_1)(w_3-w_2)}{(w_4-w_2)(w_3-w_1)} , \quad
  n = \frac{w_3-w_2}{w_3-w_1},
  \label{eq:caseIIa}
\end{equation}
where root ordering $w_1 < w_2 < w_3 < w_4$ implies
\begin{equation}
  \label{eq:n_m_ordering_caseIIa}
  0 < n < m < 1 .
\end{equation}
The physical parameters $(k,a,\w)$ are the wavenumber
\begin{equation}
  \label{eq:k_caseII}
  k = \frac{\pi}{K(m)}\sqrt{\frac{3|\beta|}{2 c^2}(w_4-w_2)(w_3-w_1)},
\end{equation}
the non-negative amplitude $a = w_3 - w_2$, and the mean strain
\begin{equation}
  \w = w_1+\dfrac{\Pi(n,m)}{K(m)}(w_2-w_1).
  \label{eq:caseIIa_mean}
\end{equation}
Examples of orbits computed using \eqref{eq:caseIIa} are shown in
Fig.~\ref{fig:orbits1}(b).

In the limit $w_3 \to w_2$, $m \to 0$, and
\begin{equation*}
  w(\theta) = w_2 - \frac{1}{2}(w_3-w_2)\cos(\theta) + \dots, \quad
  \theta = kx - \omega_0t,
\end{equation*}
subject to the linear dispersion relation \eqref{eq:disp} with
$\w = w_2$.

When $w_3 \to w_4$, $m \to 1$ and we obtain a two-parameter family of depression solitary
waves \cite{Vainchtein24}
\begin{equation}
  \label{eq:SW_caseIIa}
  w(\xi) = \frac{w_2 - w_1 n\,
    \mathrm{tanh}^2(\xi/\ell)}{1 - n\,
    \mathrm{tanh}^2(\xi/\ell)} , \quad n = \frac{w_4-w_2}{w_4-w_1},
  \quad  \ell =
  \left ( \frac{3|\beta|}{2c^2}(w_4-w_2)(w_4-w_1) \right )^{-1/2} ,
\end{equation}
with amplitude $a = w_4 - w_2$, propagating on the background
$\w = w_4$ with velocity-amplitude relation
\begin{equation}
  \label{eq:c_SW_caseIIa}
  c^2 = c_s^2(\w)  + \frac{1}{2} \beta a^2  -
  \frac{2}{3}a(\alpha+3\beta\w).
\end{equation}
An example of the corresponding homoclinic trajectory is shown by the
cyan loop in Fig.~\ref{fig:orbits1}(b).  Using root ordering
$w_1< w_2 <w_4$, \eqref{eq:root_sum} and requiring that $c^2>0$ in
\eqref{eq:c_SW_caseIIa}, we obtain the existence conditions
\begin{equation}
  \label{eq:SW_caseIIa_existence}
 0 < a < 2 \left ( \w + \frac{\alpha}{3 \beta} \right ) , \quad \w +
 \frac{\alpha}{3\beta} > 0, \quad c_s^2(\w)  + \frac{1}{2} \beta a^2  -
  \frac{2}{3}a(\alpha+3\beta\w)>0.
\end{equation}
Note that these conditions impose an \emph{upper bound} on the
amplitude of the solitary wave, as well as upper and lower bounds on
$\w$, in contrast to the waves \eqref{eq:SW1_caseIa}, whose amplitude
can be arbitrarily large for any $\w$ but may have a strictly positive
lower bound (recall \eqref{eq:SW1_caseIa_existence}). We also remark
that the first two conditions in \eqref{eq:SW_caseIIa_existence} imply
that the solitary waves are supersonic, i.e., $c^2>c_s^2(\w)$ in
\eqref{eq:c_SW_caseIIa}, and thus the third inequality is
automatically satisfied when $c_s^2(\w) \geq 0$. However, this
inequality must be imposed when $c_s^2(\w)<0$ to ensure that $c^2>0$.

Another family of periodic traveling waves is related to the first family via the symmetry transformation \eqref{eq:sym_gen}. One can show that the two families differ by a $\pi$ shift in phase $\theta$ and have the same mean, wave number and amplitude. In the limit $w_2 \to w_1$, the second family of periodic orbits approaches the two-parameter family of elevation solitary waves that can be obtained from \eqref{eq:SW_caseIIa} using \eqref{eq:sym_gen}.

The case in which $w_2 \to w_1$ and $w_3 \to w_4$ results in heteroclinic orbits described by supersonic
kink solutions that are given by \cite{Vainchtein24}
\begin{equation}
w(\xi)=\dfrac{w_1+w_4}{2} \pm \dfrac{w_4-w_1}{2}\tanh(p\xi), \quad p=\dfrac{\sqrt{6|\beta|}(w_4-w_1)}{2|c|},
\label{eq:kink}
\end{equation}
where
\begin{equation}
w_1+w_4=-\dfrac{2\alpha}{3\beta}, \quad c^2=1-\dfrac{\alpha^2}{3\beta}+\beta\left(w_1+\dfrac{\alpha}{3\beta}\right)^2.
\label{eq:kink_more}
\end{equation}
If the plus sign is chosen in \eqref{eq:kink}, $w(\xi) \to w_4$ as
$\xi \to \infty$ and $w(\xi) \to w_1$ as $\xi \to -\infty$. With the
minus sign, we have $w(\xi) \to w_1$ as $\xi \to \infty$ and
$w(\xi) \to w_4$ as $\xi \to -\infty$. The amplitude $a=w_4-w_1$ of
the kink is the upper bound \eqref{eq:SW_caseIIa_existence} of the
solitary wave amplitude for solutions \eqref{eq:SW_caseIIa} and their
symmetric counterparts. When $(\alpha,\beta) = (0,-1)$, we have
$w_1=-w_4$, $c^2=1-w_1^2=1-w_4^2$ by \eqref{eq:kink_more}, and the
solution \eqref{eq:kink} simplifies to
\[
  w(\xi) = \pm w_4 \,\mathrm{tanh} \left ( w_4 \sqrt{\frac{6}{1-w_4^2}} \ \xi \right ).
\]
Examples of periodic orbits approaching the heteroclinic ones are
shown in Fig.~\ref{fig:orbits1}(c). The two families of periodic
solutions yield the same trajectories. Near the heteroclinic orbits,
the solution acquires the flat-top structure of a periodic
kink-antikink lattice shown in Fig.~\ref{fig:orbits1}(d).

In what follows, we consider the case $(\alpha,\beta)=(0,-1)$ without loss of generality. The periodic traveling wave solution \eqref{eq:caseIIa} can be parameterized by $\mathbf{W}$ in \eqref{eq:w_parameterization} subject to the existence
inequalities \eqref{eq:w4} and \eqref{eq:root_products}.  As before,
we can also introduce the alternative parameterization $\mathbf{A}$ 
in \eqref{eq:A_parameterization_caseIandII} involving the amplitude $a$.

The transformation
$\mathbf{W} = \mathbf{W}(\mathbf{A})$ and its Jacobian determinant are
\begin{equation}
  \label{eq:A_parameterization_transformation_caseIIa}
  \begin{split}
    w_1 &= a \Xi(n,m)\gamma_1(n,m), \quad
          w_2 = a \Xi(n,m) \gamma_2(n,m), \quad
          w_3 =  a \Xi(n,m) \gamma_3(n,m), \\
    \Xi &=\frac{1}{4n(m-n)} , \quad
              \det \left (\frac{\partial \mathbf{W}}{\partial \mathbf{A}}
              \right )
              = \frac{a^2(1-n)}{4n^2(m-n)^2} ,
  \end{split}
\end{equation}
where
\begin{equation*}
  \begin{split}
    \gamma_1 &= -n^2 + 2(1+m)n - 3m, \quad \gamma_2 = 3n^2 -
               2n(1+m) + m, \\
    \gamma_3 &= -n^2 - 2n(1-m) + m .
  \end{split}
\end{equation*}
When $a > 0$ and \eqref{eq:n_m_ordering_caseIIa} hold, the transformation
\eqref{eq:A_parameterization_transformation_caseIIa} is
invertible. Periodic traveling wave existence is subject to
\eqref{eq:n_m_ordering_caseIIa} and a real wave speed requiring
\eqref{eq:root_products}:
\begin{equation*}
  \begin{split}
    \mathbf{A} \in E_a &= \bigg \{  (n,m,a,\vb)^T \in \mathbb{R}^4
                         \,|\, 0 < n < m < 
                           1, ~ 0 < a < a_{\rm max}(n,m)\bigg \}, \\
    a_{\rm max} &= \frac{\sqrt{2}}{\Xi(n,m)P(n,m)},
  \end{split}
\end{equation*}
where
\[
  \begin{split}
    P(n,m) &=  \sqrt{\gamma_1 \gamma_2
             + \gamma_1 \gamma_3 + \gamma_2 \gamma_3-(\gamma_1 +
             \gamma_2 + \gamma_3)^2} \\
           &= \sqrt{6 n^4 - 8(1+m)n^3 + 4(2+m+2m^2)n^2 -8 m(1+m)n + 6m^2}.
  \end{split}
\]
The amplitude upper bound $a_{\rm max}(n,m)$ is an increasing function
of its arguments with
$a_{\rm max}(0,0) = a_{\rm max}(0,m) = \lim_{n\to m}a_{\rm max}(n,m) =
0$ for $0 < m < 1$ and
$\lim_{n \to 1}\lim_{m\to 1} a_{\rm max}(n,m) = 2$ so that all periodic
traveling waves are constrained to have amplitudes in the interval
$0 < a < 2$.

We can obtain other parameterizations by, for example, transforming
$a$ in $\mathbf{A}$ to the wave frequency
\begin{equation*}
  \omega = \mathrm{sgn}(c)\frac{\pi a}{K(m)}
  \sqrt{\frac{3(1-n)}{2n(m-n)}} , 
\end{equation*}
which is invertible when \eqref{eq:n_m_ordering_caseIIa} holds.
Similar transformations to $k$ and $\w$ can also be obtained.

\subsection{Case III: $\beta=0$, single family of periodic orbits}
\label{sec:case-iii:-beta=0}

\begin{figure}
\centering
\includegraphics[width=0.6\textwidth]{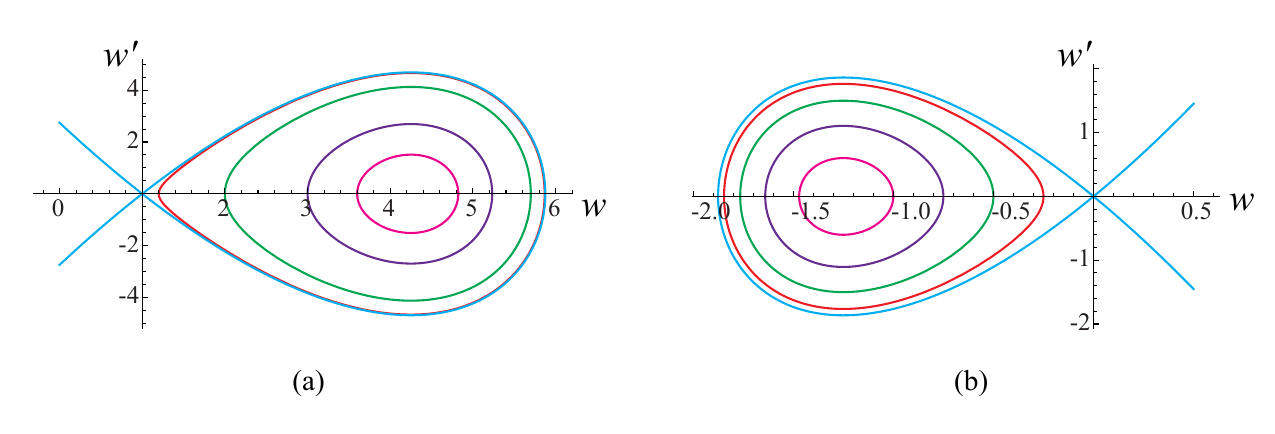}
\caption{Periodic orbits for different values of $B$ at $\alpha=1$,
  $\beta=0$, $c=2.5$, $A=-8.5$ corresponding to \eqref{eq:caseIIIa}
  with $m=0.9225$, $a=4.6671$, $\w=2.8434$ (red), $m=0.66971$,
  $a=3.6995$, $\w=3.5976$ (green), $m=0.3994$, $a=2.2333$, $\w=4.0457$
  (purple), $m=0.22765$, $a=1.2224$, $\w=4.1915$ (magenta). Manifolds
  associated with the saddle point (including the homoclinic orbit)
  are shown in cyan.}
\label{fig:orbits2}
\end{figure}

In this case, eq.~\eqref{eq:ODE_cubic} has a quadratic polynomial in
the right hand side, and we focus on the case of two real roots,
corresponding to two equilibrium points $(w_*,0)$ in the phase
plane. If $\alpha>0$, there is a saddle point on the left, and a
center on the right, so that the homoclinic trajectory enclosing the
periodic orbits corresponds to a tensile (bright) solitary wave (see
Fig.~\ref{fig:orbits2}). If $\alpha<0$, the center is on the left,
and the saddle is on the right so that the homoclinic trajectory
corresponds to a compressive (dark) solitary wave. Recall that, in this case, the periodic
orbits satisfy \eqref{eq:orbit_quad}.
For $\alpha>0$, we have $w_2 \leq w \leq w_3$, and the solution is given by \cite{Kamchatnov00}
\begin{equation}
w(\theta)=w_3-(w_3-w_2)\text{sn}^2\left(\frac{K(m)}{\pi}\theta\right), \quad m=\dfrac{w_3-w_2}{w_3-w_1},
\label{eq:caseIIIa}
\end{equation}
with amplitude $a=w_3-w_2$, wavenumber
\begin{equation}
k=\dfrac{\pi}{K(m)}\sqrt{\dfrac{2|\alpha|}{c^2}(w_3-w_1)}
\label{eq:k_caseIII}
\end{equation}
and mean
\begin{equation}
\w=w_1+(w_3-w_1)\dfrac{E(m)}{K(m)},
\label{eq:caseIIIa_mean}
\end{equation}
where $E(m)$ is the complete elliptic integral of the second
kind. Example trajectories are shown in Fig.~\ref{fig:orbits2}.

The harmonic limit when $w_3 \to w_2$, and thus $m \to 0$, is described by 
\[
w(\theta) = w_2 + \frac{1}{2}(w_3-w_2)\cos(\theta) + \dots, \quad \theta = kx-\omega_0t,
\]
with $\omega_0$ corresponding to the linear dispersion relation
\eqref{eq:disp} on the background $\w = w_2$.  In the limit
$w_2 \to w_1$, $m \to 1$, and we obtain the two-parameter family of
elevation solitary waves
\begin{equation*}
w(\xi)=w_3-(w_3-w_1)\text{tanh}^2(\xi/l), \quad l=\left(\dfrac{2\alpha}{c^2}(w_3-w_1)\right)^{-1/2},
\end{equation*}
with amplitude $a=w_3-w_1$, propagating on the background $\w=w_1$ with velocity-amplitude relation
\begin{equation*}
c^2=c_s^2(\w)+\dfrac{2\alpha}{3}a,
\end{equation*}
where we used \eqref{eq:c_vs_roots_quad}. See the cyan loop in
Fig.~\ref{fig:orbits2} for an example of the corresponding homoclinic
orbit.

For $\alpha<0$, we have $w_1 \leq w \leq w_2$, and the solution can be obtained from \eqref{eq:caseIIIa} using the symmetry transformation
\begin{equation*}
w(\theta) \to -w(\theta), \quad w_3 \to -w_1, \quad w_2 \to -w_2, \quad w_1 \to -w_3,
\end{equation*}
where again the roots are renumbered to keep their nondecreasing order.
In the limit when $w_2 \to w_3$, it yields a family of depression solitary waves. 



In what follows, we assume $\alpha = 1$.  In addition to the
parameterization $\mathbf{W}$ in \eqref{eq:w_parameterization} in
terms of the roots, it is advantageous to introduce the
parameterization 
\begin{equation}
  \label{eq:abar}
  \overline{\mathbf{A}} = (m,a,\vb,\w)^T .
\end{equation}
The transformation $\mathbf{W} = \mathbf{W}(\overline{\mathbf{A}})$
and its Jacobian determinant are
\begin{equation*}
  \begin{split}
    w_1 &= \frac{m\w K(m) - aE(m)}{m K(m)}, \quad w_2 = 
          \frac{m K(m)(\w-a) - a(E(m)-K(m))}{mK(m)}, \\
    w_3 &= \frac{m\w K(m) - a(E(m)-K(m))}{mK(m)} , \quad \det\left (
          \frac{\partial \mathbf{W}}{\partial \overline{\mathbf{A}}}
          \right ) = - \frac{a}{m^2} ,
  \end{split}
\end{equation*}
demonstrating invertibility for $a > 0$ and $m > 0$.  

The positivity of the squared wave speed \eqref{eq:c_vs_roots_quad} implies
that the mean strain \eqref{eq:caseIIIa_mean} exhibits the lower bound
\begin{equation}
  \label{eq:caseIII_mean_min}
  \w > \w_{\rm min} = a \frac{3 E(m)-(2-m)K(m)}{3 m K(m)} - \frac{1}{2} .
\end{equation}
The minimum mean is a decreasing function of $m$, exhibiting the
limits $\lim_{m \to 1} \w_{\rm min} = -\frac{1}{2} - \frac{a}{3}$,
$\lim_{m \to 0} \w_{\rm min} = -\frac{1}{2} + \frac{\omega_0^2}{24}$.
Thus, the minimum mean satisfies
\begin{equation}
  \label{eq:caseIII_mean_min_bounds}
  -\frac{a}{3} \le \w_{\rm min} +\frac{1}{2} \le \frac{\omega_0^2}{24} .
\end{equation}
The periodic traveling wave existence criteria are
\begin{equation}
  \label{eq:caseIII_existence}
  \overline{\mathbf{A}} \in \overline{E}_a = \bigg \{ (m,a,\vb,\w)^T
  \in \mathbb{R}^4 \, | \, 0 \le m \le 1, ~ a > 0, ~ \w > \w_{\rm min} \bigg \} .
\end{equation}
It first attains the value $\w_{\rm min} = -\frac{1}{2}$ when
$m = m_* \approx 0.96115$ satisfying $3 E(m_*)-(2-m_*)K(m_*)$.  That
is, for $0 < m < m_*$, the periodic traveling wave existence criteria
include $\w > \w_{\rm min}(m,a) > -\frac{1}{2}$.  When
$m_* \le m \le 1$,
$\w > \w_{\rm min}(m,a) \ge -\frac{1}{2} - \frac{a}{3}$, i.e., the
mean can fall below the value $-\frac{1}{2}$.  It will be convenient
for later analysis to express the admissible set $\overline{E}_a$ as
the union of two (not disjoint) subsets
\begin{equation}
  \label{eq:caseIII_existence_partition}
  \begin{split}
    \overline{E}_a &= \overline{E}^{(a)}_a \cup \overline{E}^{(b)}_a ,
    \\
    \overline{E}^{(\rm a)}_a &= \bigg \{ \overline{\mathbf{A}} \in
                           \mathbb{R}^4 \, | \, 0 < m < 1, ~ 0 < a <
                           a_{\rm max}(m,\w), ~ \w > -\frac{1}{2} 
                           \bigg \} , \\
    \overline{E}^{(\rm b)}_a &= \bigg \{ \overline{\mathbf{A}} \in
                           \mathbb{R}^4 \, | \, m_* \approx 0.96115 <
                           m < 1, ~ 0 < a , 
                           ~ \w_{\rm min}(m,a) < \w 
                           \bigg \} . 
  \end{split}
\end{equation}
Here, the upper bound on the amplitude of the periodic traveling wave
is
\begin{equation}
  \label{eq:caseIII_amax}
  a_{\rm max}(m,\w) =
  \begin{cases}
      3 \left ( \frac{1}{2} + \w \right ) \frac{m K(m)}{3
    E(m) - (2-m)K(m)}, & 0 < m < m_* , \\
    \infty, & m_* < m < 1 .
  \end{cases}
\end{equation}
Since
\begin{equation}
  \label{eq:caseIII_omega_transformation}
  \omega = \mathrm{sgn}(c)\frac{\pi\sqrt{2a/m}}{K(m)}, 
\end{equation}
we can invertibly transform $a \to \omega$ in \eqref{eq:abar} within the
admissible region.

When $a \to a_{\rm max}$ for $0 < m < m_*$, the periodic traveling wave
velocity $c \to 0$ but the frequency $\omega$ remains finite so that the
wavenumber $k \to \infty$.  Consequently, this is the short-wavelength
limit.

\section{Whitham Modulation Equations}
\label{sec:whith-modul-equat}

The Whitham modulation equations are obtained from an ansatz
consisting of a slowly varying periodic traveling wave
\cite{whitham_linear_1999}.  The modulation equations describe the
projection of the PDE dynamics of \eqref{eq:Bous} onto the manifold of
periodic traveling waves constrained by a quasi-linear system of
first-order PDEs of hydrodynamic type
\cite{dubrovin_hydrodynamics_1989}.

There are several, equivalent approaches to obtain the modulation
equations.  Perhaps the simplest is the average conservation law
approach described in Sec.~\ref{sec:aver-cons-laws}.  Insertion of the
appropriate periodic traveling wave family (one of
eqs.~\eqref{eq:caseIa}, \eqref{eq:caseIIa}, \eqref{eq:caseIIIa}) into
eq.~\eqref{eq:whitham_avg_cons_laws} yields the Whitham modulation
equations directly.  The benefit of this approach is that the
conservation form is obtained immediately.  One drawback to this form
of the modulation equations is that a direct computation of the
solitary-wave limit is singular because $k \to 0$ and one of the
averaged conservation laws, e.g., \eqref{eq:WCL3}, is equivalent to
\eqref{eq:WCL1} and \eqref{eq:WCL2} in this limit.  Consequently, one
equation for the modulations is not determined.

Another, mathematically appealing approach to the derivation of the
Whitham modulation equations is the method of multiple scales.  A
suitable ansatz and solvability conditions result in the modulation
equations \cite{El16}.

Yet another approach to obtain the Whitham modulation equations is the
averaged variational principle \cite{whitham_linear_1999}.
Integrating the Lagrangian $\mathbb{L}(\theta)$ in
eq.~\eqref{eq:Lagrangian} over a period of the traveling wave family
leads to the average Lagrangian
$\mathcal{L}(\mathbf{r}) = \frac{1}{2\pi} \int_0^{2\pi}
\mathbb{L}(\theta)\,\mathrm{d}\theta$, where
$\mathbf{r} \in \mathbb{R}^4$ are the periodic traveling wave's
parameters.  The Euler-Lagrange equations for the averaged Lagrangian
are the Whitham modulation equations.  One benefit of this approach is
that an additional conservation law, the conservation of wave action
\[
  \partial_T \mathcal{L}_\omega - \partial_X \mathcal{L}_k = 0,
\]
persists in the solitary-wave limit $k \to 0$.  In what follows, we
use the averaged Lagrangian method to obtain the Whitham modulation
equations.

\subsection{Averaged Conservation Laws Revisited}
\label{sec:aver-cons-laws-1}

Recall that the averaged conservation laws are given by \eqref{eq:whitham_avg_cons_laws}. Averaging of a $2\pi$-periodic function $F(\theta)$ can now be
expressed as
\[
  \overline{F} = \frac{1}{2\pi}\int_0^{2\pi}F(\theta)\,\mathrm{d}\theta =
  \frac{|c|k}{2\pi\sqrt{6|\beta|}} \oint
  \frac{F(w)}{\sqrt{|G_4(w)|}} \,\mathrm{d}w,
\]
in the case of the quartic potential $\phi(w)$.  For example, taking
three roots, say $w_1$, $w_2$, $w_3$ incorporating the constraint
\eqref{eq:root_sum}, and the averaged particle velocity $\vb$ in
\eqref{eq:v_TW} as the modulation parameters
$\mathbf{W} = (w_1,w_2,w_3,\vb)$,
eqs.~\eqref{eq:whitham_avg_cons_laws} are four quasi-linear modulation
equations for $\mathbf{W}$.  The first conservation law
\eqref{eq:WCL1} is $\w_T - \vb_X = 0$,
where $\w$ is defined in terms of the roots $w_j$ through
\eqref{eq:c_vs_roots} and one of \eqref{eq:caseIa_mean} or
\eqref{eq:caseIIa_mean} or \eqref{eq:caseIIIa_mean}, respectively.
Using $\omega = ck$, eq.~\eqref{eq:c_vs_roots} and one of
\eqref{eq:k_omega}, \eqref{eq:k_caseII}, \eqref{eq:k_caseIII} for $k$,
the conservation of waves \eqref{eq:WCL4} gives another modulation
equation.  The average velocity \eqref{eq:WCL2} and energy
\eqref{eq:WCL3} equations require further calculation of the density
and fluxes.

\subsection{Average Lagrangian}
\label{sec:average-lagrangian}

Writing the Lagrangian density \eqref{eq:Lagrangian} in terms of $u(x,t)$, we obtain
\begin{equation}
  \mathbb{L}(u_t,u_x,u_{xt}) = \dfrac{1}{2}u_t^2 - \phi(u_x) +
  \dfrac{1}{24} u_{xt}^2, \quad \phi'(u_x)=f(u_x)
  \label{eq:Lagrangian_u}
\end{equation}
and note that it depends only on derivatives of $u(x,t)$.  Furthermore, the periodic
traveling wave solutions are obtained for $v = u_t$ and $w = u_x$.
Consequently, the most general periodic traveling wave can be written as
\begin{equation}
  \label{eq:PTW_gen1}
  u(x,t) = \w x + \vb t + U(\theta), \quad U(\theta + 2\pi) =
  U(\theta), \quad  \theta = kx - \omega t .
\end{equation}
The term $\w x + \vb t$ corresponds to the so-called pseudo-phase and
enables $U(\theta)$ to be $2\pi$-periodic.



In the general case \eqref{eq:cubic}, the Lagrangian density
\eqref{eq:Lagrangian_u} is
\begin{equation}
\mathbb{L}=\dfrac{1}{2}u_t^2-\dfrac{1}{2}u_x^2-\dfrac{1}{3}\alpha
u_x^3-\dfrac{1}{4}\beta u_x^4+\dfrac{1}{24}u_{xt}^2.
\label{eq:L_quad1}
\end{equation} Note that \eqref{eq:PTW_gen1} implies that
$u_t=\vb-c(w(\theta)-\w)$, $u_x=w(\theta)$ and
$u_{xt}=-\omega w'(\theta)$, so that \eqref{eq:L_quad1} can be written
solely in terms of $w = w(\theta)$
\[
\mathbb{L}=\dfrac{1}{2}\left(\vb-c(w-\w)\right)^2-\dfrac{1}{2}w^2-\dfrac{\alpha}{3}w^3-\dfrac{\beta}{4}
w^4+\dfrac{1}{24}\omega^2(w')^2.
\]
Recalling that, in view of \eqref{eq:orbit_cubic_gen}, we have
\begin{equation}
\dfrac{\omega^2(w')^2}{24}=-\dfrac{\beta}{4}w^4-\dfrac{\alpha}{3}w^3+\dfrac{1}{2}(c^2-1)w^2+\dfrac{1}{2}Aw+\dfrac{1}{2}B,
\label{eq:wprime_quad}
\end{equation}
we arrive at
\[
\mathbb{L}=\dfrac{1}{12}\omega^2(w')^2-\left(c^2\w+\vb c+\dfrac{1}{2}A\right)w+\dfrac{1}{2}\vb^2+\dfrac{1}{2}c^2\w^2+c\,\vb \,\w-\dfrac{1}{2}B.
\]
Thus the averaged Lagrangian density is given by
\begin{equation*}
  \mathcal{L}(\w,\vb,k,\omega,A,B) = \frac{1}{2\pi} \int_0^{2\pi}
  \mathbb{L}(\theta)\,\mathrm{d}\theta
  =\dfrac{\omega^2}{12k}\mathcal{W}\left(\dfrac{\omega}{k},A,B\right)+\dfrac{1}{2}\vb^2-\dfrac{\omega^2}{2k^2}\w^2-\dfrac{1}{2}A\w-\dfrac{1}{2}B,
\end{equation*}
where
\begin{equation}
\begin{split}
\mathcal{W}(c,A,B)&=\dfrac{k}{2\pi}\int_0^{2\pi}(w'(\theta))^2d\theta=\dfrac{1}{2\pi}\oint kw'(\theta)dw\\
&=\dfrac{1}{2\pi}\oint \left[\dfrac{12}{c^2}\left(-\dfrac{\beta}{2}w^4-\dfrac{2\alpha}{3}w^3+(c^2-1)w^2+Aw+B\right)\right]^{1/2}dw.
\end{split}
\label{eq:define_W}
\end{equation}
Here we used \eqref{eq:wprime_quad}, and the integration in $w$ is around the periodic orbit.

The variational equations along with the consistency relations are \cite{whitham_linear_1999}
\begin{equation}
\begin{split}
&\mathcal{L}_A=0, \quad \mathcal{L}_B=0,\\
&\dfrac{\partial}{\partial T}\mathcal{L}_\omega-\dfrac{\partial}{\partial X}\mathcal{L}_k=0, \quad k_T+\omega_X=0,\\
&\dfrac{\partial}{\partial T}\mathcal{L}_{\vb}+\dfrac{\partial}{\partial X}\mathcal{L}_{\w}=0, \quad \w_T-\vb_X=0.
\end{split}
\label{eq:var_eqns}
\end{equation}
The first equation, $\mathcal{L}_A=0$, yields
\[
\w=\dfrac{c\omega\mathcal{W}_A}{6},
\]
which is automatically satisfied because, by \eqref{eq:orbit_cubic} or
\eqref{eq:orbit_quad} and \eqref{eq:define_W}, we have
\[
\mathcal{W}_A=\begin{cases}\dfrac{6|c|}{2\pi c^2\sqrt{6|\beta|}}\oint\dfrac{w \ dw}{\sqrt{|G_4(w)|}}, & \beta \neq 0 \\
  \dfrac{6|c|}{2\pi c^2\sqrt{8|\alpha|}}\oint\dfrac{w \ dw}{\sqrt{|G_3(w)|}}, & \beta=0,
\end{cases}
\]
which in either case yields
\[
\mathcal{W}_A=\dfrac{6}{2\pi c^2k}\int_0^{2\pi} wd\theta=\dfrac{6}{c\omega}\w.
\]
The second equation in \eqref{eq:var_eqns}, $\mathcal{L}_B=0$, implies that
\begin{equation*}
k=\dfrac{6}{\mathcal{W}_B c^2},
\end{equation*}
which yields the dispersion relation
\begin{equation*}
\omega^2=(ck)^2=\dfrac{6}{k \mathcal{W}_B}.
\end{equation*}
We also have the identity
\begin{equation*}
  \mathcal{W}_c = -\frac{1}{c} \mathcal{W} + \frac{12}{\omega}
  \overline{w^2} .
\end{equation*}

The remaining four equations in \eqref{eq:var_eqns} yield
\begin{subequations}
  \label{eq:whitham_avg_cons_laws_var}
  \begin{align}
    \label{eq:WCL1var}
    \w_T - \overline{v}_X &= 0, \\
    \label{eq:WCL2var}
    \overline{v}_T - (c^2\w+\dfrac{1}{2}A)_X &= 0, \\
    \label{eq:WCL3var}
    \left(\dfrac{c}{12}\mathcal{W} + c\dfrac{\overline{w^2} - \w^2}{k}
    \right)_T
    +\left(c^2\dfrac{\overline{w^2} -
    \w^2}{k}\right )_X &= 0 , \\
    \label{eq:WCL4var}
    k_T + \omega_X &= 0 .
  \end{align}
\end{subequations}
Here, \eqref{eq:WCL1var} and \eqref{eq:WCL4var} coincide with \eqref{eq:WCL1} and \eqref{eq:WCL4}, respectively, while \eqref{eq:WCL2var} is equivalent to \eqref{eq:WCL2} because, by \eqref{eq:ODE}, we have
\[
c^2\w+\dfrac{1}{2}A=\overline{f(w)+\dfrac{\omega^2}{12}w''(\theta)}=\overline{f(w)}.
\]
Equation \eqref{eq:WCL3var} is the conservation of wave action, which
makes \eqref{eq:whitham_avg_cons_laws_var} consistent with
\eqref{eq:whitham_avg_cons_laws} but, as we will see, remains distinct
from the remaining equations in the limit $k \to 0$, whereas
eq.~\eqref{eq:WCL3} becomes redundant with the remaining modulation
equations.

Two special limits are of interest. In the \emph{solitary-wave limit}, we have $k \to 0$, and hence $\omega \to 0$, so that \eqref{eq:WCL4var} is trivially satisfied. Thus, in this limit, the modulation system consists of three equations  \eqref{eq:WCL1var}, \eqref{eq:WCL2var} and \eqref{eq:WCL3var}. In the \emph{harmonic limit}, we have $\overline{w^2} \to \w^2$ and $\mathcal{W} \to 0$, while $k$ is nonzero, making the third equation \eqref{eq:WCL3var} trivial. Thus, the modulation system \eqref{eq:whitham_avg_cons_laws_var} in this limit reduces to three equations, \eqref{eq:WCL1var}, \eqref{eq:WCL2var} and \eqref{eq:WCL4var}. These equations are given by
\begin{equation}
\begin{split}
\w_T-\overline{v}_X&=0,\\
\overline{v}_T-c_s^2(\w)\w_X&=0,\\
\omega_T+\mathrm{sgn}(\omega)\dfrac{(c_s^2(\w)-\omega^2/12)^{3/2}}{c_s^2(\w)}\omega_X 
-\dfrac{(\alpha+3\beta\w)\omega}{c_s^2(\w)}\overline{v}_X&=0,
\end{split}
\label{eq:ME_Hlimit}
\end{equation}
where we recall that $c_s(\w)$ is the sound speed defined in \eqref{eq:sonic}.
We recognize the first two equations as the dispersionless limit of \eqref{eq:system}, as expected.  The third equation is found to be
equivalent to the modulation equation for linear waves---conservation
of waves \eqref{eq:WCL4var} with $\omega = \omega_0(k,\w)$, the linear
dispersion relation \eqref{eq:disp}, which can be inverted to yield
\[
k(\omega,\w)=\dfrac{|\omega|}{\sqrt{c_s^2(\w)-\omega^2/12}},
\]
where we note that $\omega^2<12c_s^2(\w)$ by \eqref{eq:disp}. The coefficient of
$\omega_X$ in this equation equals the group velocity
$\mathrm{sgn}(\omega) /k_\omega$, while the coefficient of
$\overline{v}_X$ is $k_{\w}/k_\omega$. In both the solitary wave and
harmonic limits, the first two mean equations, \eqref{eq:WCL1var} and
\eqref{eq:WCL2var}, decouple from the third.

We will parameterize the Whitham equations
\eqref{eq:whitham_avg_cons_laws_var} using a convenient set of
modulation variables, i.e., one of $\mathbf{W}$ in
\eqref{eq:w_parameterization}, $\mathbf{A}$ in
\eqref{eq:A_parameterization_caseIandII}, $\mathbf{\tilde{A}}$ in
\eqref{eq:A_param_caseI_complex}, or $\overline{\mathbf{A}}$ in
\eqref{eq:abar}, depending on the situation.  Note that
\eqref{eq:whitham_avg_cons_laws_var} have the form
$\mathbf{P}_T+\mathbf{Q}_X=0$, where $\mathbf{P}$ and $\mathbf{Q}$ are
the density and flux vectors, respectively.  For each parametrization
$\R$, we can find $\mathbf{P}(\R)$ and $\mathbf{Q}(\R)$ and rewrite
the Whitham modulation equations
\eqref{eq:whitham_avg_cons_laws_var} in the hydrodynamic form
\begin{equation}
  \label{eq:2}
  \R_T + \A(\R)\R_X = \mathbf{0} ,
\end{equation}
where
$\mathbb{A}(\R) = (\partial \mathbf{P}/\partial \R)^{-1} \partial
\mathbf{Q}/\partial \R$ is obtained using symbolic algebra in
Mathematica.


The modulation system $\R_T + \A(\R)\R_X = \mathbf{0}$ has four eigenvalue-eigenvector pairs
$\{(\lambda_j,\mathbf{z}_j)\}_{j=1}^4$ satisfying
\begin{equation*}
  \A(\R) \mathbf{z}_j(\R) = \lambda_j(\R) \mathbf{z}_j(\R), \quad j = 1, 2,3, 4.
\end{equation*}
We order the eigenvalues first by their real parts and, for complex
conjugate eigenvalues, then by their imaginary parts. We say that the
modulation system is \emph{strictly hyperbolic} if
$\lambda_j \le \lambda_{j+1} \in \mathbb{R}$ and
$\lambda_j = \lambda_{j+1}$ if and only if the $j^{\rm th}$ and
$(j+1)^{\rm th}$ modulation equation are equivalent. Here we use an
extended definition of strict hyperbolicity (see, e.g.,
\cite{levermore_hyperbolic_1988}) in which merged characteristic
velocities correspond to a reduction in order of the modulation
system.  For example, this is precisely the case in the solitary wave
and harmonic limits of \eqref{eq:whitham_avg_cons_laws_var}, where
either the conservation of waves \eqref{eq:WCL4var} or conservation of
wave action \eqref{eq:WCL3var} are identically zero in the respective
limits. Both limits thus maintain strict hyperbolicity of the
modulation system provided that $c_s^2(\w)=f'(\w)>0$.  The $j^{\rm th}$
characteristic field is said to be \emph{genuinely nonlinear} if
\cite{lax_hyperbolic_1973}
\begin{equation}
  \label{eq:mu_j}
  \mu_j(\R) = \nabla_{\R} \lambda_j(\R) \cdot \mathbf{z}_j(\R) \ne 0 , \quad j = 1,2,3,4.
\end{equation}
Together, strict hyperbolicity and genuine nonlinearity characterize the \emph{convexity} of the system.

In particular, the system of the first two mean equations in
\eqref{eq:ME_Hlimit}, which decouples from the remaining equation in
the harmonic and solitary-wave limits, has eigenvalues
$\lambda_{1,2}=\mp c_s(\w)$ and the corresponding eigenvectors
$\mathbf{z}_{1,2}=(1,\pm c_s(\w))^T$, yielding
$\mu_{1,2}=\mp f''(\w)/(2\sqrt{f'(\w)})$. This subsystem is strictly
hyperbolic as long as $c_s^2(\w)=f'(\w)>0$ and it is genuinely
nonlinear when $f''(\w) \neq 0$.

Considering now the harmonic-limit system \eqref{eq:ME_Hlimit}, we find that the eigenvalues are
\[
\lambda_1=-c_s(\w), \quad \lambda_2= \mathrm{sgn}(\omega)\dfrac{(c_s^2(\w)-\omega^2/12)^{3/2}}{c_s^2(\w)}, \quad \lambda_3=c_s(\w),
\]
and the corresponding eigenvectors are
\[
\begin{split}
&\mathbf{z}_1=\left(1,c_s(\w),\dfrac{(\alpha+3\beta\w)c_s(\w)\omega}{\mathrm{sgn}(\omega)
  (c_s^2(\w)-\omega^2/12)^{3/2}+c_s^3}\right)^T,
\quad \mathbf{z}_2=(0,0,1)^T, \\
&\mathbf{z}_3=\left(1,-c_s(\w),-\dfrac{(\alpha+3\beta\w)c_s(\w)\omega}{\mathrm{sgn}(\omega)(c_s^2(\w)-\omega^2/12)^{3/2}-c_s^3}\right)^T,
\end{split}
\]
yielding
\[
\mu_1=-\dfrac{f''(\w)}{2\sqrt{f'(\w)}}, \quad \mu_2=-\dfrac{(f'(\w)-\omega^2/12)^{1/2}}{4f'(\w)}|\omega|,
\quad \mu_3=\dfrac{f''(\w)}{2\sqrt{f'(\w)}} .
\]
Since the harmonic wave frequency is bounded from above
\eqref{eq:disp}, the system is strictly hyperbolic when
$c_s^2(\w) = f'(\w) >0$ and genuinely nonlinear when $f''(\w) \neq 0$
and $\omega \neq 0$. In particular, in the case $(\alpha,\beta)=(1,0)$
considered below, the harmonic limit is convex (strictly hyperbolic
and genuinely nonlinear) when $\w>-1/2$ and $\omega \neq 0$. For cubic nonlinearity with
$(\alpha,\beta)=(0,1)$, the harmonic limit system is always strictly
hyperbolic, while in the case $(\alpha,\beta)=(0,-1)$ strict
hyperbolicity requires $|\w|<1/\sqrt{3}$. In both cases, we must have
$\omega \neq 0$ and $\w \neq 0$ for genuine nonlinearity.

\subsection{Quadratic nonlinearity:  $(\alpha,\beta) = (1,0)$}
\label{sec:case-iii-alpha1beta0}

We now consider the special case of $(\alpha,\beta) = (1,0)$.  Note
that, if $\beta = 0$, we can set $\alpha = 1$ without loss of
generality using the transformation $\alpha u \to u$, as noted in
Sec.~\ref{sec:setup}.  We wish to express the Whitham
eqs.~\eqref{eq:whitham_avg_cons_laws_var} in terms of the parameter
set $\mathbf{W} = (w_1,w_2,w_3,\overline{v})^T$.  For this, we will
need to calculate
\[
\begin{split}
  \mathcal{W} &= \frac{1}{2\pi} \sqrt{\frac{8}{c^2}} \oint
  \sqrt{-G_3(w)}\,\mathrm{d} w = \frac{1}{\pi}\sqrt{\frac{8}{c^2}}
                \int_{w_2}^{w_3} \sqrt{-G_3(w)}\,\mathrm{d} w, \\
  \overline{w^2} &= \frac{1}{2\pi} \int_0^{2\pi}
                   w^2(\theta)\,\mathrm{d}\theta =  \frac{1}{2\pi} \sqrt{\frac{\omega^2}{8}} \oint
                   \frac{w^2}{\sqrt{-G_3(w)}}\,\mathrm{d}w =
                   \frac{1}{\pi} \sqrt{\frac{\omega^2}{8}}
                   \int_{w_2}^{w_3}
                   \frac{w^2}{\sqrt{-G_3(w)}}\,\mathrm{d}w .
\end{split}
\]
Evaluating the elliptic integrals, we obtain \cite{Byrd13}
\begin{equation*}
  \mathcal{W} = \frac{4\sqrt{2}}{15\pi |c|}
                \frac{(w_3-w_2)(w_2-w_1)^2}{\sqrt{w_3-w_1}}
                \frac{2(1-m+m^2) E(m) -
                (2-m)(1-m)K(m)}{m(1-m)^2}
\end{equation*}
and
\begin{equation*}
\begin{split}
  \overline{w^2} &= \frac{|\omega|}{3\sqrt{2}\pi(m-1)^2\sqrt{w_3-w_1}}
                   \{2(w_2-w_1)\left ( (1-2m)w_1+(2-m)w_2 \right )E(m)\\
                   &+ (m-1)\left((3m-2)w_1^2-2 w_1w_2 + w_2^2 \right)K(m)\}.
\end{split}
\end{equation*}

Using equations \eqref{eq:c_vs_roots_quad}, \eqref{eq:A_quad}, \eqref{eq:caseIIIa}, \eqref{eq:k_caseIII}, \eqref{eq:caseIIIa_mean} with $\alpha=1$, we obtain
\beq
\begin{split}
  A &= - \frac{2}{3}(w_1 w_2 + w_1 w_3 + w_2 w_3), \\
  c^2 &= 1 + \frac{2}{3}(w_1+w_2+w_3) , \quad \omega = c k, \\
  k &= \dfrac{\sqrt{2}\pi}{|c|K(m)}\sqrt{w_3-w_1}, \quad m =
      \frac{w_3-w_2}{w_3-w_1} , \\
  \w &= w_1+(w_3-w_1)\dfrac{E(m)}{K(m)}.
\end{split}
\label{eq:caseIII_params}
\eeq
Consequently, we have expressed each density and flux of the Whitham
equations \eqref{eq:whitham_avg_cons_laws_var} in terms of the
parameters $\mathbf{W}$ in \eqref{eq:w_parameterization}.

\subsubsection{Solitary-Wave Limit}
\label{sec:solitary-wave-limit}

We now pass to the solitary-wave limit $w_2 \to w_1$ in the Whitham
equations \eqref{eq:whitham_avg_cons_laws_var}.  The terms in the
wave-action density eq.~\eqref{eq:WCL3var} exhibit the limits
\[
\begin{split}
  \lim_{w_2 \to w_1} \frac{c}{12} \mathcal{W}
  &= \frac{\sqrt{2}\mathrm{sgn}(c)}{2\pi} \frac{4}{45} (w_3-w_1)^{5/2}, \\
  \lim_{w_2 \to w_1} c \frac{\overline{w^2} - \w^2}{k}
  &= \frac{\sqrt{2}\mathrm{sgn}(c)}{2\pi} \frac{2}{3} c^2(w_3 - w_1)^{3/2} , \\
  \lim_{w_2 \to w_1} \w &= w_1 , \\
  \lim_{w_2 \to w_1} \overline{w^2} &= w_1^2 , \\
  \lim_{w_2 \to w_1} c^2 &= 1 + \frac{2}{3}(2 w_1 + w_3).
\end{split}
\]
Since the solitary wave amplitude is $a = w_3 - w_1$
(cf.~Sec.~\ref{sec:case-iii:-beta=0}), we can express the above
quantities in terms of $\overline{\mathbf{A}} =(1,a,\vb,\w)^T$ to obtain
\begin{subequations}
  \label{eq:whitham_caseIIIa_soliton}
  \begin{align}
    \label{eq:ME1}
    \w_T - \overline{v}_X &= 0 , \\
    \label{eq:ME2}
    \overline{v}_T - f(\w)_X &= 0 , \\
    \label{eq:ME3}
    \left ((1+2\w)a^{3/2} + \tfrac{4}{5} a^{5/2} \right
    )_T + \left ( c \left ( (1+2\w)a^{3/2} +
    \tfrac{2}{3} a^{5/2} \right ) \right )_X &= 0 ,
  \end{align}
  where
  \begin{equation*}
    c^2 = 1 + 2\w + \frac{2}{3} a , \quad f(\w) = \w + \w^2 .
  \end{equation*}
\end{subequations}
Note that in eq.~\eqref{eq:ME3}, we have multiplied through by the
inconsequential constant $2\pi/(\sqrt{2}\mathrm{sgn}(c))$ to simplify
the expression.  We recognize eqs.~\eqref{eq:ME1} and \eqref{eq:ME2} as
the dispersionless limit of eq.~\eqref{eq:system} and
eq.~\eqref{eq:ME3} is the equation for the solitary-wave amplitude
field $a = a(X,T)$.  Expanding eq.~\eqref{eq:ME3} and replacing $\w_T =
\vb_X$, we obtain
\begin{equation}
  \label{eq:AE}
  a_T +  c \,a_X + \frac{4 a}{3(1+2\w+\frac{4}{3}a)} \vb_X +
  \frac{2 c a}{1+2\w+\frac{4}{3}a} \w_X
   = 0 .
\end{equation}
Because it is decoupled from the other equations \eqref{eq:ME1},
\eqref{eq:ME2}, the characteristic velocity for \eqref{eq:AE} is
$\lambda=c$, as it should be in the solitary-wave limit.

Equation \eqref{eq:whitham_caseIIIa_soliton} exhibits the ordered
characteristic velocities (for the branch of fast waves with $c > 0$)
\begin{equation*}
  \lambda_1 = -\sqrt{1+2\w}, \quad \lambda_2 = \sqrt{1+2\w}, \quad
  \lambda_3 = c = \sqrt{1 + 2\w + \tfrac23 a} .
\end{equation*}
The corresponding eigenvectors are
\begin{equation*}
  \mathbf{z}_1 =
  \begin{bmatrix}
    1 \\ \sqrt{1+2\w} \\ -
    \frac{1+2\w+2a+\sqrt{(1+2\w)(1+2\w+\frac{2}{3}a)}}{1+2 \w +
    \frac{4}{3}a} 
  \end{bmatrix}, \quad
  \mathbf{z}_2 =
  \begin{bmatrix}
    1 \\ \sqrt{1+2\w} \\ -
    \frac{1+2\w+2a-\sqrt{(1+2\w)(1+2\w+\frac{2}{3}a)}}{1+2 \w +
    \frac{4}{3}a} 
  \end{bmatrix}, \quad
  \mathbf{z}_3 =
  \begin{bmatrix}
    0 \\ 0 \\ 1
  \end{bmatrix} .
\end{equation*}

\subsubsection{Convexity}
\label{sec:hyperbolicity}

We first examine the convexity (hyperbolicity, genuine nonlinearity)
of the Whitham equations in the solitary-wave limit
\eqref{eq:whitham_caseIIIa_soliton}.  The mean lower bound
$\w > -\frac{1}{2}-\frac{a}{3}$ for solitary waves
(recall \eqref{eq:caseIII_mean_min_bounds}) ensures that
$\lambda_3 \in \mathbb{R}$ but it does not preclude the possibility of
complex characteristic velocities $\lambda_1$ and $\lambda_2$.  Thus,
to ensure strict hyperbolicity, we further require $\w > -\frac{1}{2}$
and $a > 0$.  To assess genuine nonlinearity, we compute
\begin{equation*}
\begin{split}  
    \mu_j &= \nabla \lambda_j \cdot \mathbf{z}_j = \partial_{\w}
            \lambda_j = \frac{1}{\lambda_j} \ne 0, \quad j = 1,2, \\
    \mu_3 &= \nabla \lambda_3 \cdot \mathbf{z}_3 = \partial_a
            \lambda_3(a,\w) = \frac{1}{3
            \lambda_3} \ne 0 .
\end{split}
\end{equation*}
Thus the solitary-wave modulation system is genuinely nonlinear,
provided it is strictly hyperbolic.

We now examine the convexity of the full Whitham system
\eqref{eq:whitham_avg_cons_laws_var} by numerically computing the
eigenvalues and eigenvectors of the modulation matrix $\A(\R)$ in
\eqref{eq:2} as $\overline{\mathbf{A}}$ is varied.  Recalling the
admissibility criteria \eqref{eq:caseIII_existence},
\eqref{eq:caseIII_existence_partition}, we will study convexity in two
separate regimes:
\begin{enumerate}
\item[(a)] $\overline{E}_a^{(\rm a)}$: $\w > -\frac{1}{2}$,
  $0 < a < a_{\rm max}$, and $0 < m < 1$,
\item[(b)] $\overline{E}_a^{(\rm b)}$: $\w_{\rm min}(m,a) < \w$,
  $a > 0$, and $m_* < m < 1$,
\end{enumerate}
where $\w_{\rm min}$ is given by  \eqref{eq:caseIII_mean_min} and $a_{\rm max}$ is defined in \eqref{eq:caseIII_amax}.
For case (a), we introduce the change of variables
\begin{equation*}
  \mathrm{(a)} \quad u \to (1+2\w) u + \vb t + \w x, \quad t \to \frac{t}{\sqrt{1+2\w}},
  \quad \w > -\frac{1}{2},
\end{equation*}
to eq.~\eqref{eq:Bous}, leaving it unchanged since $\w >
-\tfrac12$. Consequently, in this case, we will set $\vb = \w = 0$
without loss of generality and consider variation of the remaining two
modulation parameters $0 < m < 1$, $0 < a < a_{\rm max}(m,\w = 0)$.
For case (b), we note that eq.~\eqref{eq:Bous} is invariant under the alternative change of variables
\begin{equation*}
  \mathrm{(b)} \quad u \to \frac{u}{a} + \vb t + \left ( \frac{1}{2a} - \frac{1}{2} \right
  ) x, \quad t \to \sqrt{a} t,
  \quad a > 0,
\end{equation*}
Now, we will set
$a = 1$ without loss of generality and consider convexity when
$m_* < m < 1$ and $\w > \w_{\rm min}(m,a=1)$.


\begin{figure}
  \centering
  \includegraphics[scale=0.22]{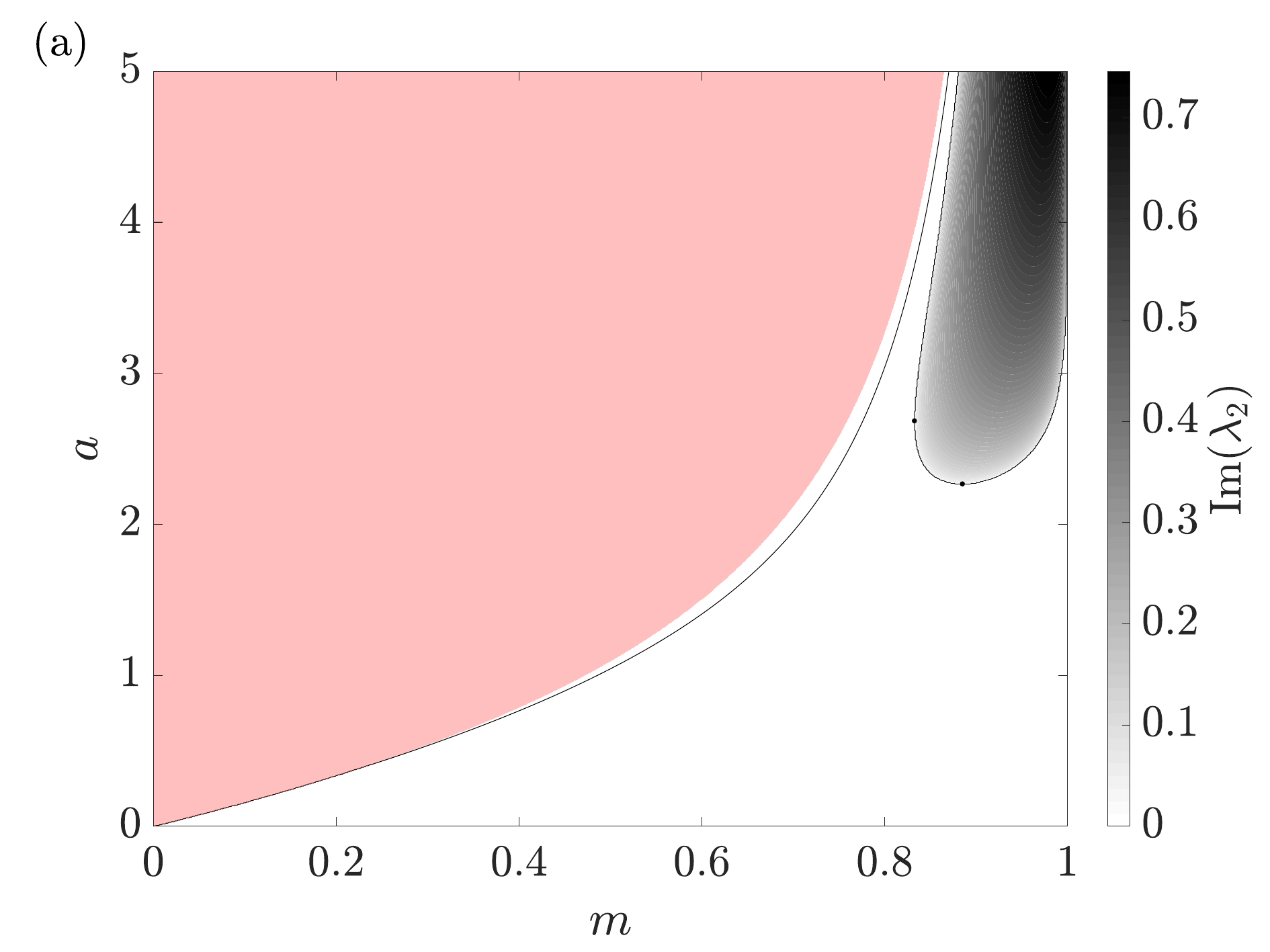} ~
  \includegraphics[scale=0.22]{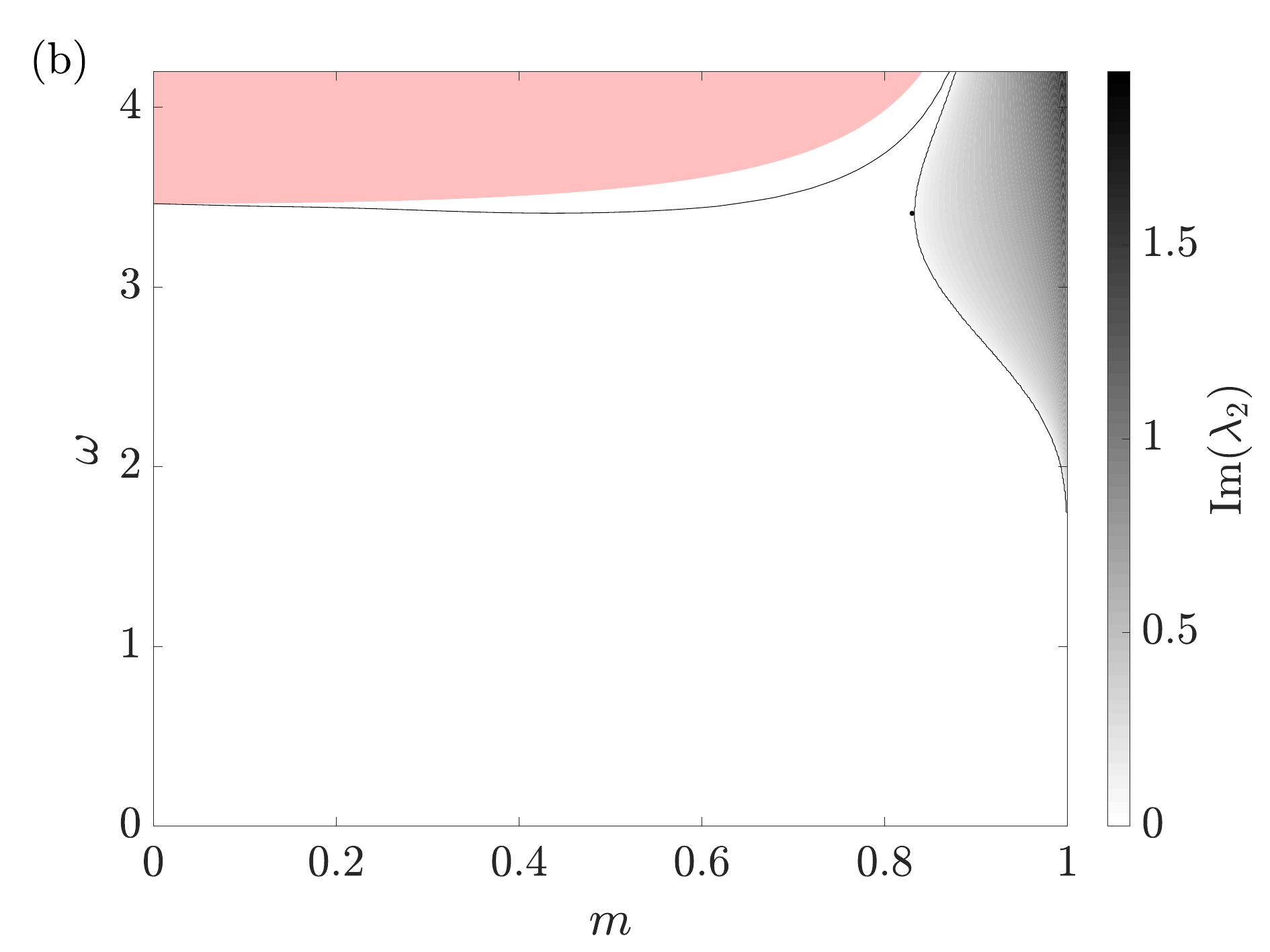}
  \caption{Hyperbolicity of the regularized Boussinesq-Whitham
    equations for $(\alpha,\beta) = (1,0)$ (case III), quadratic
    interaction force with
    $\overline{\mathbf{A}} = (m,a,\vb=0,\w=0) \in
    \overline{E}_a^{(a)}$: Strict hyperbolicity (white
    region). Grayscale is the magnitude of the imaginary part of
    $\lambda_2$ where the system is not hyperbolic. Periodic traveling
    waves do not exist in the inadmissible pink region.  Black dots
    depict points corresponding to the values
    $(m,a) \in \{(0.83,2.69),~(0.89,2.27)\}$ in (a) and
    $(m,\omega) = (0.83,3.41)$ in (b) where complex characteristic
    velocities are first encountered.  The solid black curves correspond
    to parameters where the second characteristic field loses genuine
    nonlinearity $\mu_2 = \nabla \lambda_2 \cdot \mathbf{z}_2 =
    0$. (a) and (b) are equivalent under the transformation
    \eqref{eq:caseIII_omega_transformation}}
  \label{fig:hyperbolicityIII}
\end{figure}

Figure \ref{fig:hyperbolicityIII} depicts the (white) region of strict
hyperbolicity when
$\overline{\mathbf{A}} \in \overline{E}_a^{(\rm a)}$.  We find that
$\lambda_3 < \lambda_4$ are real and distinct for all computed
admissible parameter values. There exists a grayscale region in which
the modulations exhibit complex conjugate characteristic velocities
$\lambda_1$ and $\lambda_2$.  Such regions correspond to modulational
instability of the underlying periodic traveling wave
\cite{whitham_linear_1999}.  At the boundaries of these regions,
$\lambda_1 = \lambda_2 \in \mathbb{R}$.  The coalescence of two
characteristics coincides with loss of genuine nonlinearity, so
$\mu_1 = \mu_2 = 0$ also. The smallest values of $m$ and $a$ where
complex characteristics occur are depicted as black dots.

We also compute the quantities $\mu_j$ in \eqref{eq:mu_j} in the strictly
hyperbolic region and find that only $\mu_2 = 0$ on a curve
bifurcating from $(a,m) = (0,0)$ or $(\omega,m) = (\sqrt{12},0)$.  The
curve is depicted in Fig.~\ref{fig:hyperbolicityIII}.  Nonzero $\mu_j$
is a necessary monotonicity condition for the existence of a
$j$-integral curve in which $\R'(\alpha) = \mathbf{z}_j(\R(\alpha))$
where $\alpha$ evolves along the characteristic
$\frac{\mathrm{d}x}{\mathrm{d}t} = \lambda_j$:
$\alpha_t + \lambda_j(\R(\alpha)) \alpha_x = 0$.  Simple waves are
used in the solution of the Riemann problem consisting of step initial
conditions at the origin.
\begin{figure}
  \centering
  \includegraphics[scale=0.25]{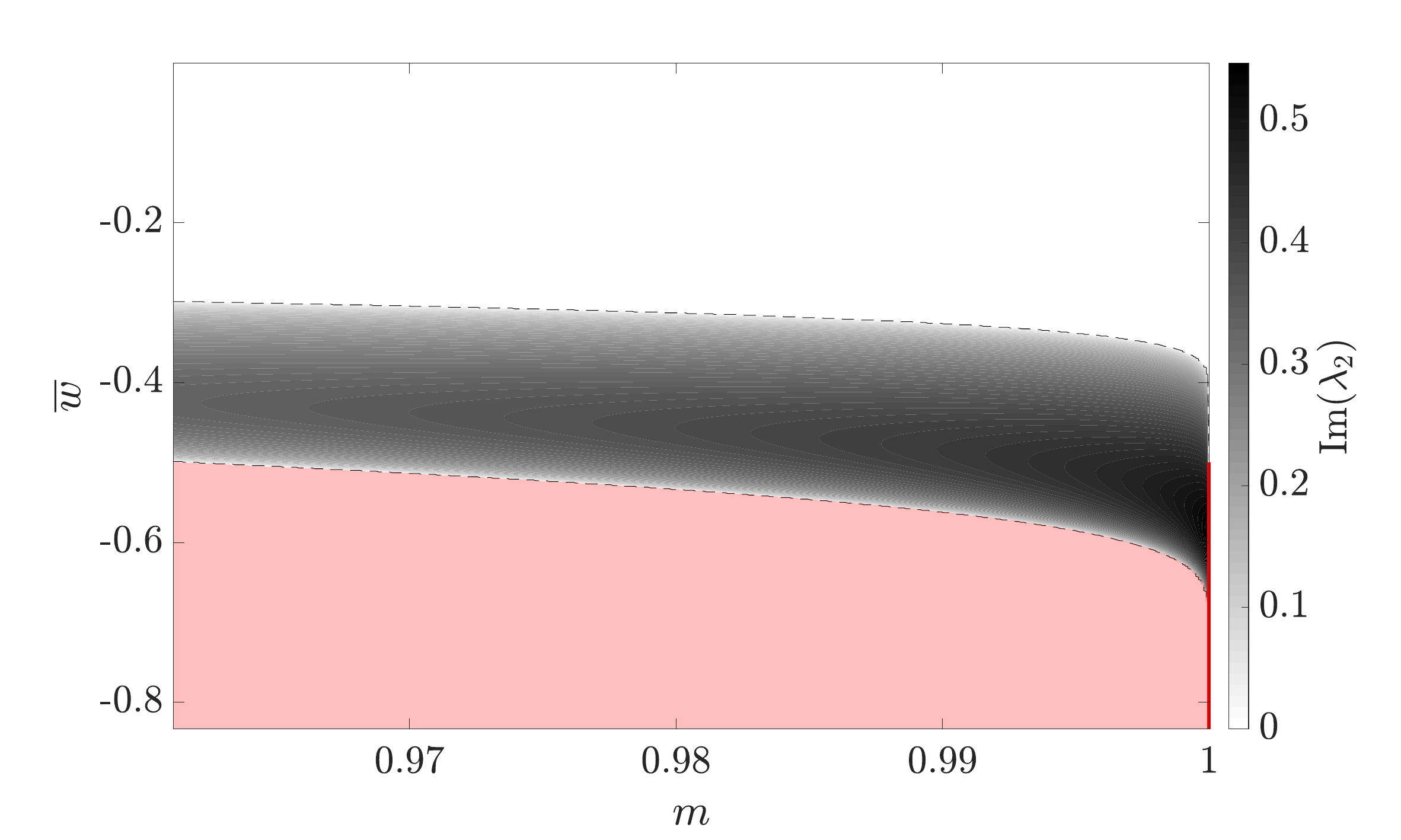}
  \caption{Hyperbolicity of the regularized Boussinesq-Whitham
    equations for case III, quadratic interaction force with
    $\overline{\mathbf{A}} = (m,a=1,\vb=0,\w) \in
    \overline{E}_a^{(b)}$: Strict hyperbolicity (white
    region). Grayscale is the magnitude of the imaginary part of
    $\lambda_2$ where the system is not hyperbolic.  At
    $m = m_* \approx 0.96115$, we find that for $\w > -0.298$, the
    system is strictly hyperbolic. The dashed curves correspond to the
    boundary of the non-hyperbolic region where
    $\lambda_1 = \lambda_2$, so that $\mu_1 = \mu_2 = 0$ there
    also. Periodic traveling waves do not exist in the inadmissible
    pink region.  The red segment at $m = 1$ corresponds to the region
    of non-hyperbolicity of the solitary wave limiting system
    \eqref{eq:whitham_caseIIIa_soliton}}
  \label{fig:hyperbolicityIII_wbar}
\end{figure}
The short-wave limit ($k \to \infty$) occurs when $a$ approaches the
pink admissibility boundary ($a \to a_{\rm max}$).  At this boundary,
the modulation matrix $\A$ exhibits a double zero eigenvalue and two
real, oppositely signed eigenvalues.

Figure \ref{fig:hyperbolicityIII_wbar} depicts hyperbolicity for case
(b) where we normalize the amplitude to $a = 1$ and require
$\w > \w_{\rm min}$, $m_* \le m < 1$.  We observe that the system is
hyperbolic when $\w$ is sufficiently large.  The depicted results are
consistent with the solitary wave limit
\eqref{eq:whitham_caseIIIa_soliton} that is strictly hyperbolic for
$\w > -\frac{1}{2}$.  The upper dashed curve in
Fig.~\ref{fig:hyperbolicityIII_wbar} where $\lambda_1 = \lambda_2$
limits to $\w = -\frac{1}{2}$ as $m \to 1$ but the approach is very
sharp due to the logarithmic approach of $\w$ to $-\frac{1}{2}$
because $K(m) \sim -1/\log(1-m)$ in the mean equation
\eqref{eq:caseIII_params} as $m \to 1$.  In a similar manner, the
lower dashed curve in Fig.~\ref{fig:hyperbolicityIII_wbar} at the
boundary of the admissible region also sharply limits to
$\w = \w_{\rm min}(a=1,m=1) = -\frac{5}{6}$ as $m \to 1$.  This
demonstrates that the region in Fig.~\ref{fig:hyperbolicityIII_wbar}
where hyperbolicity is lost bifurcates from the segment
$-\frac{5}{6} < \w < -\frac{1}{2}$ for $m = 1$, which is denoted by
the red segment.

\subsection{Cubic nonlinearity with $(\alpha,\beta) = (0,1)$}
\label{sec:case-i-alpha0beta1}
We now consider the case of cubic $f(w)$ with $(\alpha,\beta)=(0,1)$,
to which, as we discussed in Sec.~\ref{sec:setup}, case I in
Sec.~\ref{sec:case-i-pos-beta} can be reduced upon a proper
rescaling. As before, we need to compute $\mathcal{W}$ and
$\overline{w^2}$ that are given by
\[
\mathcal{W} = \frac{1}{2\pi} \sqrt{\frac{6}{c^2}} \oint \sqrt{-G_4(w)}\, \mathrm{d} w, \quad
\overline{w^2} = \frac{1}{2\pi} \sqrt{\frac{\omega^2}{6}} \oint \dfrac{w^2}{\sqrt{-G_4(w)}}\, \mathrm{d} w.
\]
Recall that in this case we have two families of periodic solutions related by the symmetry transformation \eqref{eq:sym_gen} with $\alpha=0$ and $\beta=1$. Thus, it suffices to consider the first solution family \eqref{eq:caseIa}. We have
\begin{equation}
\begin{split}
&w_1 + w_2 + w_3 + w_4 = 0,\\
& A = \frac{1}{2}(w_1w_2w_3 + w_1 w_2 w_4+w_1 w_3 w_4+w_2 w_3
      w_4),\\
&c^2 = 1 - \frac{1}{2} ( w_1 w_2 + w_1 w_3 + w_1 w_4 + w_2 w_3 +
        w_2 w_4 + w_3 w_4 ),\\
&k = \frac{\pi}{K(m)} \sqrt{\frac{3}{2c^2}(w_4-w_2)(w_3-w_1)}, \quad \omega=ck,\\
&m = \frac{(w_4-w_3)(w_2-w_1)}{(w_4-w_2)(w_3-w_1)},
\end{split}
\label{eq:caseI_param}
\end{equation}
where we used \eqref{eq:root_sum}, \eqref{eq:c_vs_roots}, \eqref{eq:A}, \eqref{eq:caseIa} and \eqref{eq:k_omega}.




To find $\mathcal{W}$ and $\overline{w^2}$ we use
\[
\begin{split}
\mathcal{W} &= \frac{1}{\pi} \sqrt{\frac{6}{c^2}} \int_{w_1}^{w_2} \sqrt{(w-w_1)(w_2-w)(w_3-w)(w_4-w)}\, \mathrm{d} w, \\
\overline{w^2} &= \frac{1}{\pi} \sqrt{\frac{\omega^2}{6}} \int_{w_1}^{w_2} \dfrac{w^2\,\mathrm{d} w}{\sqrt{(w-w_1)(w_2-w)(w_3-w)(w_4-w)}}.
\end{split}
\]
Evaluation of the elliptic integrals yields \cite{Byrd13}
\begin{equation*}
\begin{split}
\mathcal{W}&=\dfrac{(w_2 - w_1)^2 (w_3-w_1)^{1/2} (w_4 -w_1)^2}{4|c| \sqrt{6}\pi
    n^3 (1 + n)^2 (n +
     m)^2  (w_4-w_2)^{3/2}} \bigg\{n (3 n^4 + 3 m^2  + 4 n^3 (1 + m)\\
     &+ 4 n m (1 + m) + 2 n^2 (2 + m + 2 m^2)) E(m) \\
        &- (n + m) (3 n^4 - 4 n^2 (m-1) - 3 m^2 -
        6 n m^2 + 2 n^3 (2 + m)) K(m) \\
        &+ 3 (n^2 - m) (2 n + n^2 + m) (n^2 + m +
        2 n m) \Pi(-n, m))\bigg\}
\end{split}
\end{equation*}
and
\begin{equation}
\begin{split}
\overline{w^2}&=\dfrac{\sqrt{2}w_4^2|\omega|}{\sqrt{3}\pi \sqrt{(w_3 - w_1) (w_4 - w_2)}}\bigg\{K(m) + 2\left(\dfrac{w_1}{w_4}-1\right)\Pi(-n,m)\\
&+ \dfrac{(w_4 -w_1)^2}{2 (1 +n) (n + m) w_4^2} \bigg(n E(m) - (n + m) K(m) \\
&+ (n^2 + 3 m +
            2 n (1 + m)) \Pi(-n, m)\bigg)\bigg\},
\end{split}
\label{aver_wsq_caseIa}
\end{equation}
where we recall that $m$ and $n$ are given in \eqref{eq:caseIa} and
satisfy \eqref{eq:nm_bounds_caseI_real} and
\eqref{eq:n_bounds_caseI_complex} in the case of real and complex
$w_{3,4}$, respectively.

\subsubsection{Solitary-wave limit}
Recall that in the solitary-wave limit $w_2 \to w_3$ ($m \to 1$), the solutions approach the two-parameter family \eqref{eq:SW1_caseIa} with amplitude $a=w_3-w_1$, background $\w=w_3$ and velocity \eqref{eq:c_SW1_caseIa}. In this limit we obtain
\[
\begin{split}
\lim_{w_2 \to w_3}\dfrac{c}{12}\mathcal{W}&=\dfrac{\text{sgn}(c)}{\pi\sqrt{6}}\bigg\{\dfrac{1}{6}\sqrt{a(a-4\w)}(a^2 - 4 a \w + 6 \w^2)\\
&+(a -2 \w)^2 \w \arctan\sqrt{\dfrac{a}{a - 4\w}}\biggl\},\\
\lim_{w_2 \to w_3}c\dfrac{\overline{w^2}-\w^2}{k}&=\dfrac{c^2\text{sgn}(c)}{\pi\sqrt{6}} \bigg\{\sqrt{a(a-4\w)}+4\w\arctan\sqrt{\dfrac{a}{a - 4\w}}\bigg\},\\
\lim_{w_2 \to w_3}\overline{w^2} &= \w^2, \quad \lim_{w_2 \to w_3}c^2=1+3\w^2+\dfrac{1}{2}a(a-4\w),
\end{split}
\]
where we recall that either $\w<0$ and $a>0$ or $a>4 \w > 0$ by \eqref{eq:SW1_caseIa_existence} with $\alpha=0$.
In this limit, the Whitham modulation equations \eqref{eq:whitham_avg_cons_laws_var} with $f(w)=w+w^3$ become
\begin{equation}
\begin{split}
\w_T-\overline{v}_X&=0,\\
\overline{v}_T-(1+3\w^2)\w_X&=0,\\
a_T+c\,a_X+c\,\dfrac{g(a,\w)}{d(a,\w)}\w_X+\dfrac{h(a,\w)}{d(a,\w)}\overline{v}_X&=0,
\end{split}
\label{eq:ME_SWlimit_caseI}
\end{equation}
where
\begin{equation}
\begin{split}
g(a,\w)&=-\sqrt{a(a-4\w)}(2 + 4 a^2 - 19 a \w + 24 \w^2)\\
&+2 (2 + (a - 6 \w) (a - 4 \w)) (a - 2 \w)\arctan\sqrt{\dfrac{a}{a - 4 \w}},\\
h(a,\w)&=-\sqrt{a(a-4\w)} (2 + 4 a^2 - 19 a \w + 24 \w^2)\\
&+ (4 + 3 (a - 4 \w)^2) (a - 2 \w) \arctan\sqrt{\dfrac{a}{a - 4 \w}},\\
d(a,\w)&=\sqrt{a(a-4\w)} (1 + 2 a^2 - 8 a \w + 9 \w^2) +6  (a - 2 \w)^2 \w \arctan\sqrt{\dfrac{a}{a - 4 \w}}.
\end{split}
\label{eq:ME_SWlimit_caseIa}
\end{equation}
The ordered characteristic velocities (for the fast waves with $c>0$)
are
\begin{equation}
\label{eq:char_speeds_caseIa}
\lambda_1 = -\sqrt{1+3\w^2}, \quad \lambda_2 = \sqrt{1+3\w^2}, \quad
\lambda_3 = c =
\sqrt{1+3\w^2+\dfrac{1}{2}a(a-4\w)} .
\end{equation}
The associated eigenvectors are
\begin{equation}
  \label{eq:evecs}
  \mathbf{z}_j =
  \begin{bmatrix}
    1 \\ -\lambda_j \\ \frac{\lambda_j h(a,\w) - \lambda_3
    g(a,\w)}{d(a,\w)(\lambda_3 - \lambda_j)}
  \end{bmatrix}
  , \quad j = 1,2, \quad \mathbf{z}_3 =
  \begin{bmatrix}
    0 \\ 0 \\ 1
  \end{bmatrix} .
\end{equation}

In the limit $w_2 \to w_3 \to w_4$ the solitary waves reduce to the
one-parameter family \eqref{eq:SW2_caseIa} with background
$\w=w_4$, amplitude $a=w_4-w_1=4\w$, where the
second equality follows from the first equation in
\eqref{eq:caseI_param}, and velocity \eqref{eq:c_SW2_caseIa}, yielding
\[
\begin{split}
\lim_{w_2 \to w_3 \to w_4}\dfrac{c}{12}\mathcal{W}&=\dfrac{\text{sgn}(c)}{32\sqrt{6}}a^3, \quad
\lim_{w_2 \to w_3 \to w_4}c\dfrac{\overline{w^2}-\w^2}{k}=\dfrac{c^2\text{sgn}(c)}{2\sqrt{6}}a,\\
\lim_{w_2 \to w_3 \to w_4}\overline{w^2} &= \w^2, \quad \lim_{w_2 \to w_3 \to w_4}c^2=1+\dfrac{3}{16}a^2=c_s^2(\w).
\end{split}
\]
In this case the third equation in \eqref{eq:ME_SWlimit_caseI}
decouples from the first two and becomes $a_T+c\,a_X=0$, which for
$c>0$ is equivalent to the fast characteristic family in the first two
equations of \eqref{eq:ME_SWlimit_caseI}, i.e.,
$\lambda_2 = \lambda_3$.  Thus, the algebraic solitary wave
\eqref{eq:SW2_caseIa} with $c=c_s(\w) > 0$ ($c=-c_s(\w)<0$) moves with
the fast (slow) characteristic velocity of mean evolution.

\subsubsection{Convexity}
\label{sec:convexityI}

Here we examine the convexity of the Whitham modulation equations
using various means. First, we consider convexity of the solitary-wave
modulation system \eqref{eq:ME_SWlimit_caseI} analytically and then
proceed to assess the convexity of the full modulation equations
\eqref{eq:whitham_avg_cons_laws_var} by numerically evaluating the
eigenstructure of the modulation matrix $\A$ in \eqref{eq:2}.

The solitary-wave existence conditions $\w < 0$ and $a>0$ or
$a > 4\w > 0$ in eq.~\eqref{eq:SW1_caseIa_existence} with $\alpha=0$
for the first solitary-wave family \eqref{eq:SW1_caseIa} imply real
characteristic velocities \eqref{eq:char_speeds_caseIa}.  When
$a = 4 \w$, either $\lambda_2 = \lambda_3$ (for $c > 0$) or
$\lambda_1 = c = \lambda_2 = -\sqrt{1+3\w^2}$ (for $c < 0$).  This
case corresponds to the algebraic solitary wave, whose modulation is
equivalent to the fast or slow mean modulation.
Consequently, the
solitary-wave modulation system \eqref{eq:ME_SWlimit_caseI} with \eqref{eq:ME_SWlimit_caseIa} is
strictly hyperbolic.

From \eqref{eq:char_speeds_caseIa} and \eqref{eq:evecs} it follows that
the genuine nonlinearity conditions \eqref{eq:mu_j} are
\begin{align*}
  \mu_j &= \nabla \lambda_j \cdot \mathbf{z}_j = \partial_{\w}
          \lambda_j = \frac{3
          \w}{\lambda_j} \ne 0, \quad j = 1,2, \\
  \mu_3 &= \nabla \lambda_3 \cdot \mathbf{z}_3 = \partial_a
          \lambda_3(a,\w) = \frac{a  - 2\w}{2
          \lambda_3} \ne 0 .
\end{align*}
Consequently, the solitary-wave modulation system
\eqref{eq:ME_SWlimit_caseI} with \eqref{eq:ME_SWlimit_caseIa} loses
genuine nonlinearity in both of the mean modulation characteristic
families at an inflection point of the interaction force
$f''(\w) = 3 \w = 0$.  Since either $\w < 0$ and $a > 0$ or
$a > 4 \w > 0$ for solitary wave existence, we have $\mu_3 > 0$.

Thus, the modulation system in the solitary-wave limit is strictly
hyperbolic for all parameter values and loses genuine nonlinearity
when $\w = 0$.  Recall that the same is true for the harmonic limit in
this case.  Since $\w = w_3$, $a = w_3 - w_1$, and
$w_4 = -2w_3-w_1$, the loss of genuine nonlinearity occurs precisely
when $n = (w_3-w_1)/(w_4-w_3) = 1$.

\begin{figure}
  \centering
  \includegraphics[scale=0.25]{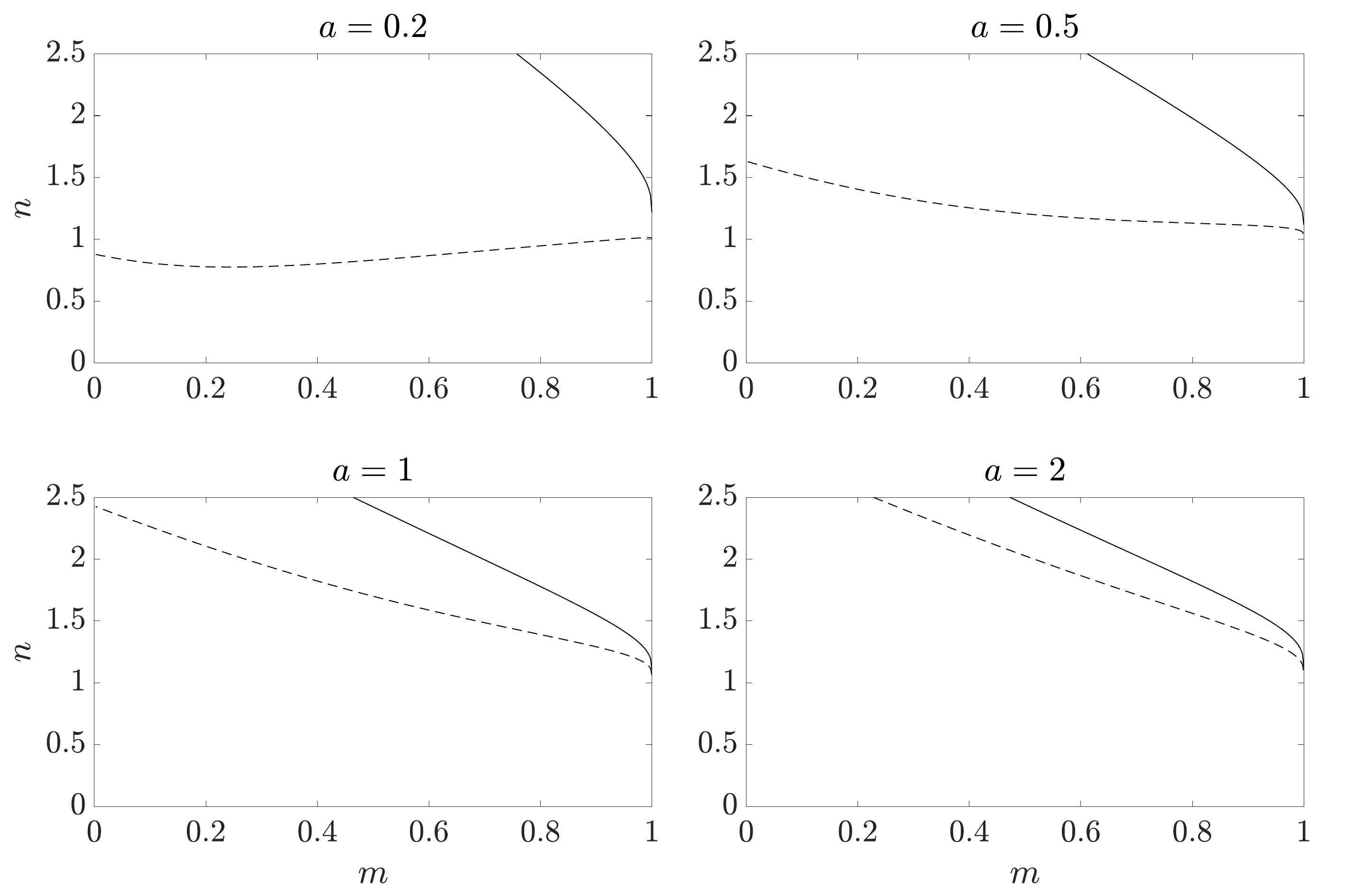}
  \caption{Convexity of the Whitham modulation equations for
    $(\alpha,\beta) = (0,1)$ (case I) with real roots $w_j$.  For
    each fixed amplitude $a$, the solid (dashed) curves correspond to
    the loss of genuine nonlinearity $\mu_1 = 0$ ($\mu_2 = 0$)
    bifurcating from the point $(n,m) = (1,1)$ where the solitary-wave
    limiting system loses genuine nonlinearity at $\w = 0$}
  \label{fig:hyperbolicityIa}
\end{figure}
For the full Whitham modulation system in the case of real roots
$w_j$, we find that the equations are strictly hyperbolic for $n>0$,
$0<m<1$.  Figure \ref{fig:hyperbolicityIa} depicts curves in the
$m$-$n$ plane with fixed $a$ where the modulation system loses genuine
nonlinearity in one of the first two characteristic families.  These
curves bifurcate from the $m = 1$ solitary-wave limit at $n = 1$, the
inflection point of the interaction force $\w = 0$ where the solitary
wave limiting system \eqref{eq:ME_SWlimit_caseI} loses genuine
nonlinearity.  Since the modulated periodic traveling wave amplitude
is held constant, the limit $m \to 0$ corresponds to the trigonometric
wave limit where $w_3 \to w_4$.

\begin{figure}
  \centering
  \includegraphics[scale=0.25]{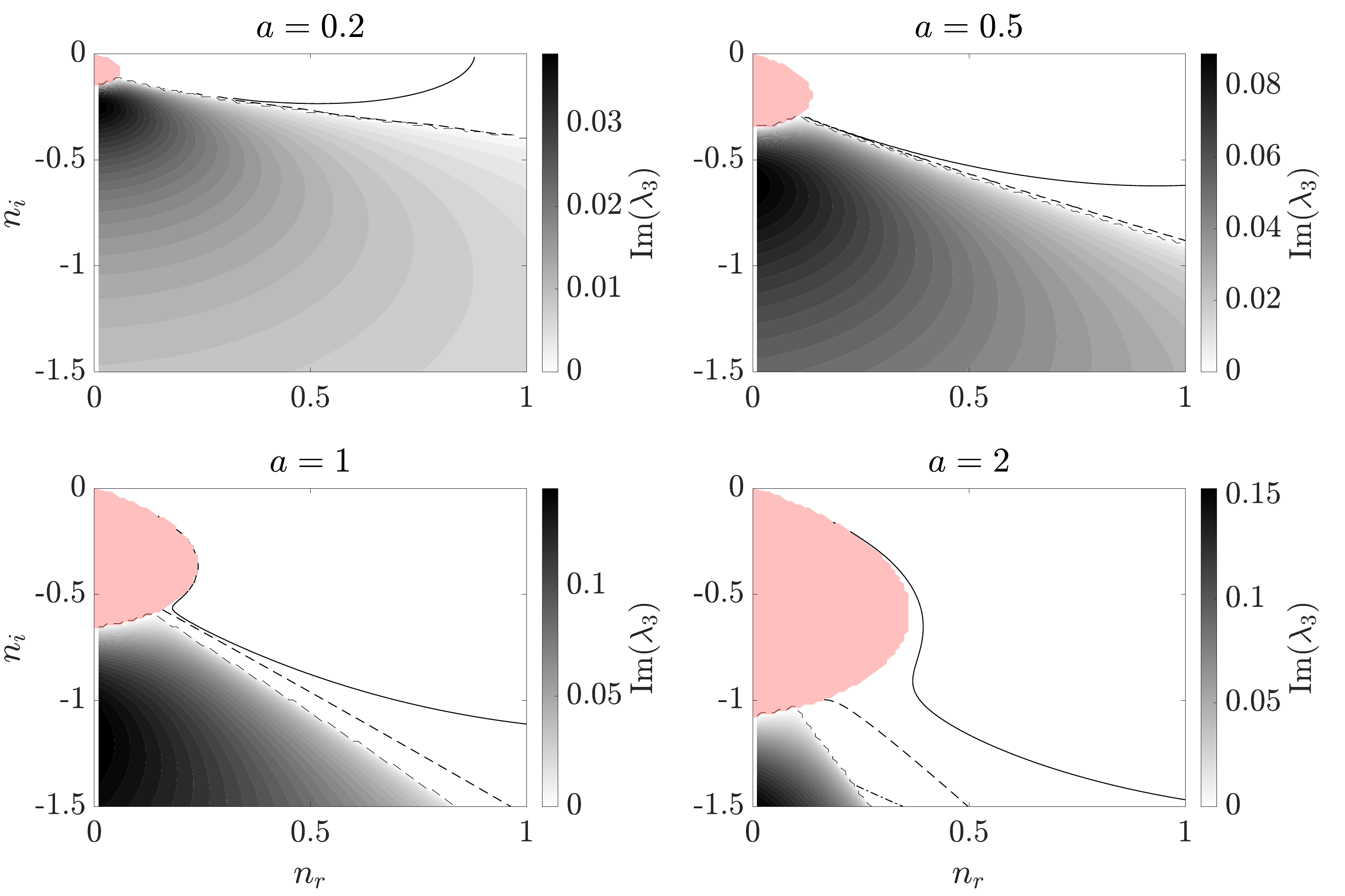}
  \caption{Convexity of the Whitham modulation equations for
    $(\alpha,\beta) = (0,1)$ (case I), cubic interaction force with
    complex conjugate roots $w_{3,4}$,
    $\tilde{\mathbf{A}} = (n_r,n_i,a,\vb=0) \in \tilde{E}_a$: The
    system is strictly hyperbolic in the white region.  Grayscale is
    $\mathrm{Im}(\lambda_3)$ where the system is not hyperbolic.
    Periodic traveling waves do not exist in the inadmissible pink
    region.  For each fixed amplitude $a$, the solid (dashed) curves
    correspond to the loss of genuine nonlinearity $\mu_2 = 0$
    ($\mu_3 = 0$).  For $a = 2$, the dash-dotted curve corresponds to
    $\mu_1 = 0$}
  \label{fig:hyperbolicityIa_complex}
\end{figure}
On the other hand, modulations of periodic traveling waves with
complex conjugate roots $w_3 = \bar{w}_4$ exhibit strict hyperbolicity in
a limited range of the $n_r$-$n_i$ plane.  This is shown in
Fig.~\ref{fig:hyperbolicityIa_complex} for differing, fixed amplitudes
$a$.  Accompanying the loss of hyperbolicity for some parameter
regions is the loss of genuine nonlinearity in the first, second, and
third characteristic fields.  The second characteristic field is
linearly degenerate along a curve $\mu_2 = 0$ that bifurcates from the
trigonometric limit $n_i = 0$.  For example, when $a = 0.2$ in
Fig.~\ref{fig:hyperbolicityIa}, the level curve $\mu_2 = 0$ meets the
trigonometric limit when $(n,m) \approx (0.878,0)$.  Similarly, in
Fig.~\ref{fig:hyperbolicityIa_complex} with $a = 0.2$, the curve
$\mu_2 = 0$ enters the domain from $(n_r,n_i) \approx (0.878,0)$.  As
$a$ varies, the level curves exhibit somewhat complex deformations.
The level curve $\mu_1 = 0$ appears within the parameter window of
consideration when $a = 2$.  The boundary of the non-hyperbolic region
is where $\lambda_2 = \lambda_3$, so $\mu_2 = \mu_3 = 0$ there.

\subsection{Cubic nonlinearity with $(\alpha,\beta) = (0,-1)$}
\label{sec:case-ii-alpha0beta-1}
Next, we discuss the case of cubic $f(w)$ with
$(\alpha,\beta)=(0,-1)$, which corresponds to case II in
Sec.~\ref{sec:case-ii-neg-beta} after an appropriate rescaling
discussed in Sec.~\ref{sec:setup}. In this setting we have
\[
\begin{split}
\mathcal{W} &= \frac{1}{2\pi} \sqrt{\frac{6}{c^2}} \oint \sqrt{G_4(w)}\, \mathrm{d} w = \frac{1}{\pi} \sqrt{\frac{6}{c^2}} \int_{w_2}^{w_3} \sqrt{(w-w_1)(w-w_2)(w_3-w)(w_4-w)}\, \mathrm{d} w, \\
\overline{w^2} &= \frac{1}{2\pi} \sqrt{\frac{\omega^2}{6}} \oint \dfrac{w^2}{\sqrt{G_4(w)}}\, \mathrm{d} w
=\frac{1}{\pi} \sqrt{\frac{\omega^2}{6}} \int_{w_2}^{w_3} \dfrac{w^2\,\mathrm{d} w}{\sqrt{(w-w_1)(w-w_2)(w_3-w)(w_4-w)}}.
\end{split}
\]
Recall that in this case there are two families of periodic solutions
related by symmetry \eqref{eq:sym_gen} with $\alpha=0$, $\beta=-1$ and
that differ by a $\pi$ phase shift, so it suffices to consider one of
them given by \eqref{eq:caseIIa}. We have
\begin{equation*}
\begin{split}
&w_1 + w_2 + w_3 + w_4 = 0,\\
& A = -\frac{1}{2}(w_1w_2w_3 + w_1 w_2 w_4+w_1 w_3 w_4+w_2 w_3
      w_4),\\
&c^2 = 1 + \frac{1}{2} ( w_1 w_2 + w_1 w_3 + w_1 w_4 + w_2 w_3 +
        w_2 w_4 + w_3 w_4 ),\\
&k = \frac{\pi}{K(m)} \sqrt{\frac{3}{2c^2}(w_4-w_2)(w_3-w_1)}, \quad \omega=ck,\\
&m = \frac{(w_4-w_1)(w_3-w_2)}{(w_4-w_2)(w_3-w_1)},
\end{split}
\end{equation*}
where we used \eqref{eq:root_sum}, \eqref{eq:c_vs_roots}, \eqref{eq:A}, \eqref{eq:caseIIa} and \eqref{eq:k_caseII}, 
and obtain \cite{Byrd13}
%
%
\begin{equation}
\begin{split}
\mathcal{W}&=\dfrac{(w_3 - w_2)^2 (w_4-w_2)^{1/2} (w_2 -w_1)^2}{4|c| \sqrt{6}\pi
    n^3 (1 - n)^2 (m-n)^2  (w_3-w_1)^{3/2}} \bigg\{n (3 n^4 + 3 m^2  - 4 n^3 (1 + m)\\
     &- 4 n m (1 + m) + 2 n^2 (2 + m + 2 m^2)) E(m) \\
        &+(m-n) (3 n^4 - 4 n^2 (m-1) - 3 m^2 +
        6 n m^2 - 2 n^3 (2 + m)) K(m) \\
        &+ 3 (m-n^2) (-2 n + n^2 + m) (n^2 + m -
        2 n m) \Pi(n, m))\bigg\}
\end{split}
\label{eq:W_caseIIa}
\end{equation}
and
\begin{equation}
\begin{split}
\overline{w^2}&=\dfrac{\sqrt{2}w_1^2|\omega|}{\sqrt{3}\pi \sqrt{(w_3 - w_1) (w_4 - w_2)}}\bigg\{K(m) + 2\left(\dfrac{w_2}{w_1}-1\right)\Pi(n,m)\\
&+ \dfrac{(w_2 -w_1)^2}{2 (n-1) (m-n) w_1^2} \bigg(n E(m) + (m-n) K(m) \\
&- (n^2 + 3 m -
            2 n (1 + m)) \Pi(n, m)\bigg)\bigg\},
\end{split}
\label{aver_wsq_caseIIa}
\end{equation}
where we recall that $m$ and $n$ are given in \eqref{eq:caseIIa} and
satisfy \eqref{eq:n_m_ordering_caseIIa}.

\subsubsection{Solitary-wave limit}
In the limit $w_3 \to w_4$ ($m \to 1$), the solutions approach the two-parameter family \eqref{eq:SW_caseIIa} of solitary waves with amplitude $a=w_4-w_2$, background $\w=w_4$ and velocity \eqref{eq:c_SW_caseIIa}. In this limit, we obtain
\begin{equation}
\begin{split}
\lim_{w_3 \to w_4}\dfrac{c}{12}\mathcal{W}&=\dfrac{\text{sgn}(c)}{\pi\sqrt{6}}\bigg\{\dfrac{1}{6}\sqrt{a(4\w-a)}(a^2 - 4 a \w + 6 \w^2)\\
&-(a -2 \w)^2 \w\, \text{arctanh}\sqrt{\dfrac{a}{4\w-a}}\biggl\},\\
\lim_{w_3 \to w_4}c\dfrac{\overline{w^2}-\w^2}{k}&=\dfrac{c^2\text{sgn}(c)}{\pi\sqrt{6}} \bigg\{-\sqrt{a(4\w-a)}+4\w\, \text{arctanh}\sqrt{\dfrac{a}{4\w-a}}\bigg\},\\
\lim_{w_3 \to w_4}\overline{w^2} &= \w^2, \quad \lim_{w_3 \to w_4}c^2=1-3\w^2+\dfrac{1}{2}a(4\w-a),
\end{split}
\label{eq:SW1_caseIIa_more}
\end{equation}
where we recall that $0<a<2\w$ by \eqref{eq:SW_caseIIa_existence} with
$\alpha = 0$. The Whitham modulation equations
\eqref{eq:whitham_avg_cons_laws_var} with $f(w)=w-w^3$ therefore
become
\begin{equation}
\begin{split}
\w_T-\overline{v}_X&=0,\\
\overline{v}_T-(1-3\w^2)\w_X&=0,\\
a_T+c\,a_X+c\,\dfrac{g(a,\w)}{d(a,\w)}\w_X+\dfrac{h(a,\w)}{d(a,\w)}\overline{v}_X&=0,
\end{split}
\label{eq:ME_SWlimit_caseII}
\end{equation}
where
\begin{equation}
\begin{split}
g(a,\w)&=\sqrt{a(4\w-a)}(-2 + 4 a^2 - 19 a \w + 24 \w^2)\\
&+2 (-2 + (a - 6 \w) (a - 4 \w)) (a - 2 \w)\,\text{arctanh}\sqrt{\dfrac{a}{4 \w-a}},\\
h(a,\w)&=\sqrt{a(4\w-a)} (-2 + 4 a^2 - 19 a \w + 24 \w^2)\\
&+(-4 + 3 (a - 4 \w)^2) (a - 2 \w)\, \text{arctanh}\sqrt{\dfrac{a}{4 \w-a}},\\
d(a,\w)&=\sqrt{a(4\w-a)} (-1 + 2 a^2 - 8 a \w + 9 \w^2) -6  (a - 2 \w)^2 \w\, \text{arctanh}\sqrt{\dfrac{a}{4 \w-a}}.
\end{split}
\label{eq:ME_SWlimit_caseIIa}
\end{equation}
The ordered characteristic velocities (for the fast-wave family with $c>0$)
are
\begin{equation}
  \label{eq:char_speeds_caseIIa}
  \lambda_1 = -\sqrt{1-3\w^2}, \quad \lambda_2 = \sqrt{1-3\w^2}, \quad
  \lambda_3 = c = \sqrt{1-3\w^2+\dfrac{1}{2}a(4\w-a)} .
\end{equation}
Note that the existence conditions \eqref{eq:SW_caseIIa_existence} for
$(\alpha,\beta)=(0,-1)$ amount to satisfying one of the following
inequality pairs
\begin{subequations}
  \label{eq:caseII_depression_soli_existence}
  \begin{align}
    \label{eq:caseII_depression_soli_existence_a}
    \mathrm{either} \quad 0 < a < 2\w \quad \mathrm{and} \quad 0 < \w
    \le \frac{1}{\sqrt{3}}, 
    \\
    \label{eq:caseII_depression_soli_existence_b}
    \mathrm{or} \quad 2\w - \sqrt{2(1-\w^2)} < a < 2\w \quad
    \mathrm{and} \quad \frac{1}{\sqrt{3}} < \w < 1 .
  \end{align}
\end{subequations}
The associated eigenvectors are given by \eqref{eq:evecs}.

\subsubsection{Kink limit}
\label{sec:kink-limit}

The limit $a \to 2\w$ corresponds to the supersonic kink solution
\eqref{eq:kink}.  To obtain its modulation, it is tempting to send
$a \to 2\w$ in eq.~\eqref{eq:ME_SWlimit_caseII}.  But this results in
a singularity.  Instead, we return to the wave-action equation
\eqref{eq:WCL3var} with the solitary-wave limiting density and flux in
eq.~\eqref{eq:SW1_caseIIa_more}.  We denote the limit as $a \to 2\wh$
in order to distinguish $\wh$ from $\w$.  An expansion of the density and flux results in
\[
  \begin{split}
    \frac{c}{12} \mathcal{W} + c \frac{\overline{w^2}-\w^2}{k}
    &= -\frac{2\mathrm{sgn}(c)}{\pi\sqrt{6}}f(\wh) \log(2\wh-a) +
                               \mathcal{O}(1) , \\
    c^2 \frac{\overline{w^2}-\w^2}{k}
    &= -\frac{2|c|}{\pi\sqrt{6}}f(\wh) \log(2\wh-a) + \mathcal{O}(1) ,
  \end{split}
\]
as $a \to 2\wh$ where $f(\wh) = \wh-\wh^3$ is the interaction force
and $c^2 = 1-\wh^2$ is the squared kink velocity \eqref{eq:kink_more}.  Inserting
this into the conservation of wave action \eqref{eq:WCL3var}, we
obtain the expansion
\[
  \frac{f(\wh)}{2\wh-a}\left [ (2\wh-a)_T + c (2\wh-a)_X \right ] +
  \log(2\wh-a) \left [ f(\wh)_T + \left ( c f(\wh) \right )_X \right ] +
  \cdots = 0 . 
\]
Multiplying this expression by $2\wh-a$ and equating each order to zero
as $a \to 2\wh$, we obtain the kink conservation law
\begin{equation*}
  f(\wh)_T + \left [ c f(\wh) \right]_X = 0, \quad f(\wh) = \wh-\wh^3,
  \quad c = \pm \sqrt{1-\wh^2} .
\end{equation*}
This equation is distinct from the mean equations for $(\w,\vb)$ in
\eqref{eq:ME_SWlimit_caseII} and admits the discontinuous traveling
wave solution
\begin{equation}
  \label{eq:undercompressive_dsw}
  \wh(X,T) = \pm 2w_0 \,\mathrm{sgn}(X-cT), \quad c^2 =
  1-w_0^2, \quad 0 < w_0 < 1  ,
\end{equation}
whose velocity $s = c$ results from the Rankine-Hugoniot jump
condition $-s[f(\wh)] + [cf(\wh)] = 0$ where $[\cdot]$ is the
difference between the state $\cdot$ to the right and left of the
discontinuity.  The weak solution \eqref{eq:undercompressive_dsw} is
the zero-dispersion limit of the kink solution \eqref{eq:kink} and
thus is an admissible shock.  Since the mean field propagates along
characteristics with velocities
$\lambda_j = \mp \sqrt{f'(\w)} = \mp \sqrt{1-3\w^2}$, they are
unchanged on either side of the shock ($f'(w_0) = f'(-w_0)$), i.e.,
the characteristics pass through the shock.  It is because of this
that the solution \eqref{eq:undercompressive_dsw} can be interpreted
as a non-classical, undercompressive shock \cite{Herrmann10,el_dispersive_2017}
and is an example of a superkink in the classification of steady
transition fronts in mechanical systems
\cite{gorbushin_transition_2022}.

\subsubsection{Convexity}
\label{sec:convexityII}
We start by considering the modulation system
\eqref{eq:ME_SWlimit_caseII} with \eqref{eq:ME_SWlimit_caseIIa} for
the solitary-wave family \eqref{eq:SW_caseIIa}, with eigenvalues
\eqref{eq:char_speeds_caseIIa} and eigenvectors \eqref{eq:evecs}. The
existence condition for this family is
\eqref{eq:caseII_depression_soli_existence}. 
In this case, we obtain
\[
\mu_j=-\dfrac{3\w}{\lambda_j},\quad j=1,2, \quad \mu_3=\dfrac{2\w-a}{2\lambda_3}.
\]
Thus, the system is strictly hyperbolic when $0 < \w < 1/\sqrt{3}$ and
genuinely nonlinear when $a \neq 2\w$, which ensures that
$ \mu_3 \neq 0$.  Since a nonzero mean is required for the existence
of a solitary wave, $\mu_j \ne 0$ for $j = 1,2$.  We recall that the
harmonic limit is strictly hyperbolic when $|\w|<1/\sqrt{3}$ and
genuinely nonlinear when $\w \neq 0$ and $\omega \neq 0$.

\begin{figure}
  \centering
  \includegraphics[scale=0.225]{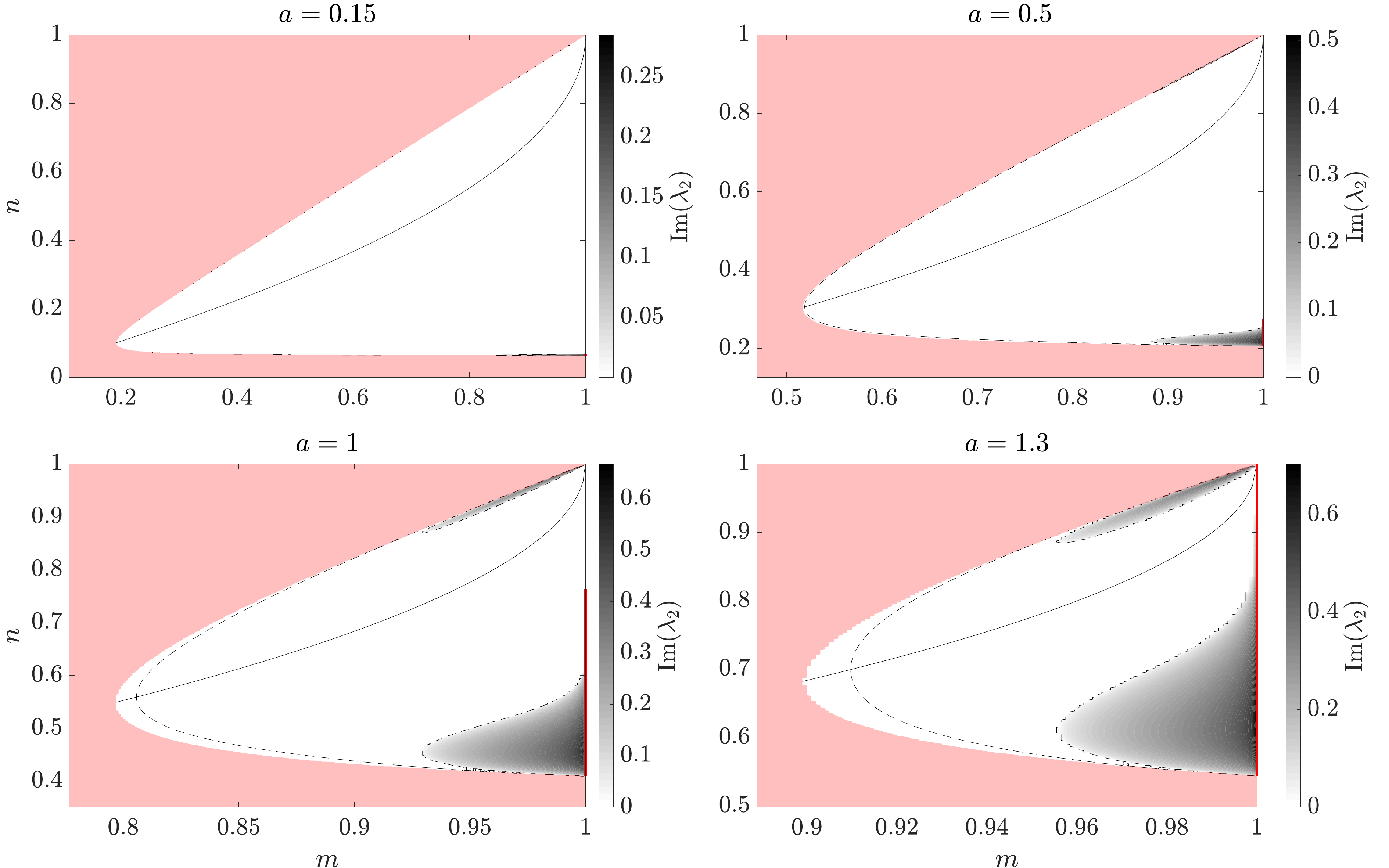}
  \caption{Hyperbolicity of the Whitham modulation equations for
    $(\alpha,\beta) = (0,-1)$ (case II).  For each fixed amplitude
    $a$, the grayscale contour plot conveys the imaginary part of a
    characteristic velocity. Periodic waves do not exist in the
    inadmissible pink region. The solid (dashed) black curves
    correspond to the loss of genuine nonlinearity $\mu_1 = \mu_4 = 0$
    ($\mu_2 = 0$).  The red segment at $m = 1$ corresponds to the
    region of non-hyperbolicity of the solitary wave limiting system
    \eqref{eq:ME_SWlimit_caseII}}
  \label{fig:hyperbolicityIIa}
\end{figure}
To investigate the convexity of the full Whitham modulation system
\eqref{eq:whitham_avg_cons_laws_var} with \eqref{eq:W_caseIIa} and
\eqref{aver_wsq_caseIIa}, we will present our results in terms of
$\mathbf{A}$ in \eqref{eq:A_parameterization_caseIandII}.  For several
choices of the periodic traveling wave amplitude $a$, we plot the
existence region and imaginary part of the second eigenvalue in
Fig.~\ref{fig:hyperbolicityIIa}.  Within the existence region, two
islands nucleate where hyperbolicity is lost.  The islands take up a
larger percentage of the existence region as the wave amplitude is
increased.  To further understand these islands, we note that the
depression solitary-wave limit with $m \to 1$ admits the following
relation between the solitary-wave parameters $(a,\w)$ and the
parameter $n$ in \eqref{eq:SW_caseIIa}
\begin{equation}
  \label{eq:caseII_n_SW}
  n = \frac{a}{4\w-a} , \quad 0 < a < 2 \w , \quad \w > 0 ,
\end{equation}
by $a = w_4 - w_2$, $\w = w_4$ and $w_1 + w_2 + 2w_4 = 0$. According
to \eqref{eq:char_speeds_caseIIa}, we expect the nucleation of
non-hyperbolic regions in the second characteristic family from
$\w \ge 1/\sqrt{3}$ in the solitary-wave limit.  Indeed, the red
segment in Fig.~\ref{fig:hyperbolicityIIa} along $m = 1$ corresponds
to the values of $n$ in \eqref{eq:caseII_n_SW} satisfying the
inequalities \eqref{eq:caseII_depression_soli_existence_b} that yield
$1/\sqrt{3} < \w < a/3 + \sqrt{2(6-a^2)}/6$.

Figure \ref{fig:hyperbolicityIIa} also depicts the level curves
$\mu_j = 0$.  The solid black curve corresponds to both $\mu_1 = 0$
and $\mu_4 = 0$.  It bifurcates from the kink limit $m \to 1$,
$n \to 1$.  The dashed curves correspond to regions where $\mu_2 = 0$.
Necessarily, these surround the non-hyperbolic grayscale regions where
$\lambda_1 = \lambda_2$ because the coalescence of two characteristics
(non-strict hyperbolicity) coincides with loss of genuine
nonlinearity, so that $\mu_1 = 0$ also there.

\section{Stability of periodic traveling waves}
\label{sec:stab}
The loss of hyperbolicity in the modulation equations discussed in
Secs.~\ref{sec:hyperbolicity}, \ref{sec:convexityI}, and
\ref{sec:convexityII} corresponds to modulational instability of the
corresponding periodic traveling wave \cite{whitham_linear_1999}. To
verify the onset of such an instability, we numerically compute the
spectrum of the linearized operator for various parameter values and
solve initial value problems with suitably perturbed periodic
traveling wave initial data.

To obtain the corresponding eigenvalue problem, we first observe that
\eqref{eq:system} is equivalent to
\begin{equation}
w_t=v_x, \qquad v_t=\left(1-\dfrac{1}{12}\partial_x^2\right)^{-1}(f(w))_x,
\label{eq:system1}
\end{equation}
where we used $w_{tt}=v_{xt}$. Rewriting \eqref{eq:system1}
in terms of $w=\tilde{w}(\xi,t)$, $v=\tilde{v}(\xi,t)$, $\xi=x-ct$, we obtain
\[
\tilde{w}_t=\tilde{v}_\xi+c\tilde{w}_\xi, \qquad \tilde{v}_t=\left(1-\dfrac{1}{12}\partial_\xi^2\right)^{-1}(f(w))_\xi+c\tilde{v}_\xi.
\]
Seeking solutions of this system in the form
\[
\left[\begin{array}{c} \tilde{w}\\\tilde{v}\end{array}\right]=\left[\begin{array}{c}W(\xi)\\V(\xi)\end{array}\right]+\eps e^{\lambda t}
\boldsymbol{\phi}(\xi),
\]
where $W(\xi)$ and $V(\xi)$ are the strain and particle velocity for a periodic traveling wave with period $T_0$, 
\[
W(\xi+T_0)=W(\xi), \quad V(\xi+T_0)=V(\xi),
\]
and keeping only terms of order $\eps$, we obtain the following
eigenvalue problem for the linearized operator:
\begin{equation}
\lambda \boldsymbol{\phi}=\left[\begin{array}{ll} c\partial_\xi & \;\;\partial_\xi\\ 
                        \left(1-\dfrac{1}{12}\partial_\xi^2\right)^{-1}\partial_\xi f'(W(\xi)) & \;\;c\partial_\xi\end{array}\right]\boldsymbol{\phi}.
\label{eq:evalue1}                        
\end{equation}
By Floquet theory applied to the $T_0$-periodic linear operator in the right hand side of \eqref{eq:evalue1}, the eigenvalue $\lambda$ is in the spectrum 
if and only if there exists a nonzero eigenfunction $\boldsymbol{\phi}$ such that \cite{Chicone99}
\[
\boldsymbol{\phi}(\xi+T_0)=e^{\frac{2\pi iq\xi}{T_0}}\boldsymbol{\phi}(\xi) 
\]
for some real $q$. This implies \cite{Bronski23} that there exists $\boldsymbol{\psi}$ such that
\begin{equation}
\boldsymbol{\phi}(\xi+T_0)=e^{\frac{2\pi i \tau \xi}{T_0}}\boldsymbol{\psi}(\xi), \quad \boldsymbol{\psi}(\xi+T_0)=\boldsymbol{\psi}(\xi),
\label{eq:psi1}
\end{equation}
where $\tau \in (-1/2,1/2]$ is the Floquet parameter. Thus, we have
\begin{equation}
\boldsymbol{\psi}(\xi)=\sum_{q\in\mathbb{Z}}\hat{\boldsymbol{\psi}}_q \exp\dfrac{2\pi i q\xi}{T_0}\,
\label{eq:psi2}
\end{equation}
where $\hat{\boldsymbol{\psi}}_q$ are the Fourier coefficients of $\boldsymbol{\psi}$.
Using \eqref{eq:evalue1} and \eqref{eq:psi1}, we obtain
\begin{equation}
\begin{split}
&\lambda \boldsymbol{\psi}=\left[\begin{array}{ll} c\left(\frac{2\pi i\tau}{T_0}+\partial_\xi\right) & \;\;\frac{2\pi i\tau}{T_0}+\partial_\xi\\ 
                        M(\tau) f'(W(\xi)) & \;\;c\left(\frac{2\pi i\tau}{T_0}+\partial_\xi\right)\end{array}\right]\boldsymbol{\psi}, \\
&M(\tau)=\left(1-\dfrac{1}{12}\left(\frac{2\pi i\tau}{T_0}+\partial_\xi\right)^2\right)^{-1}\left(\frac{2\pi i\tau}{T_0}+\partial_\xi\right).
\end{split}                        
\label{eq:evalue2}                        
\end{equation}

To numerically compute the spectrum of the linear operator, we use the
Floquet-Fourier-Hill method \cite{DeconinckKutz06}. Following the
procedure in \cite{Bronski23}, we restrict the values of $q$ in
\eqref{eq:psi2} to $q=-N_q,\dots,N_q$ and use Fourier collocation to
project \eqref{eq:evalue2} to a subspace of
$L^2(-T_0/2,T_0/2) \times L^2(-T_0/2,T_0/2)$. This reduces the
problem to finding eigenvalues of a $(4N_q+1)\times(4N_q+1)$ matrix
$\mathbf{L}(\tau)$, which is computed by taking sums and products of
diagonal and Toeplitz matrices that represent differential operators
and multiplications by periodic functions in \eqref{eq:evalue2},
respectively. Computing these eigenvalues for each value of $\tau$ in
the $N_\tau$-point discretization of $(-1/2,1/2]$, we obtain the
numerical approximation of the spectrum. In what follows, we use
$N_q=100$ and $N_\tau=200$. Since
$f'(W(\xi))=1+2\alpha W(\xi)+3\beta W^2(\xi)$, the procedure requires
computing the Fourier coefficients of $W(\xi)$ in the quadratic
nonlinearity case $(\alpha,\beta)=(1,0)$ and $W^2(\xi)$ in the case of
cubic nonlinearity, where $(\alpha,\beta)=(0,\pm 1)$. In the case of
quadratic nonlinearity, the coefficients can be found analytically
using the Fourier expansion for $\text{sn}^2(x)$ \cite{Kiper84}:
\[
W(\xi)=\sum_{q\in\mathbb{Z}} \hat{W}_q e^\frac{2\pi iq\xi}{T_0}, \quad 
\hat{W}_q=\begin{cases} (w_3-w_2)\frac{\pi^2q\kappa^q}{m K^2(m)(1-\kappa^{2q})}, & q\neq 0 \\
                        w_3-(w_3-w_2)\frac{K(m)-E(m)}{mK(m)}, & q=0
          \end{cases}, \quad \kappa=\exp\dfrac{\pi K(1-m)}{K(m)}.
\]
For the cubic nonlinearity cases, the Fourier coefficients were
computed using numerical integration in Mathematica. To speed up the
computations, we took advantage of the fact that these coefficients
quickly decay as $|q|$ tends to infinity so we restricted the computed
coefficients to $|q| \leq q_\text{cut}$, setting the rest to zero. In
most cases, $q_\text{cut}=40$ was sufficient. 

\begin{figure}
\centering
\includegraphics[width=0.8\textwidth]{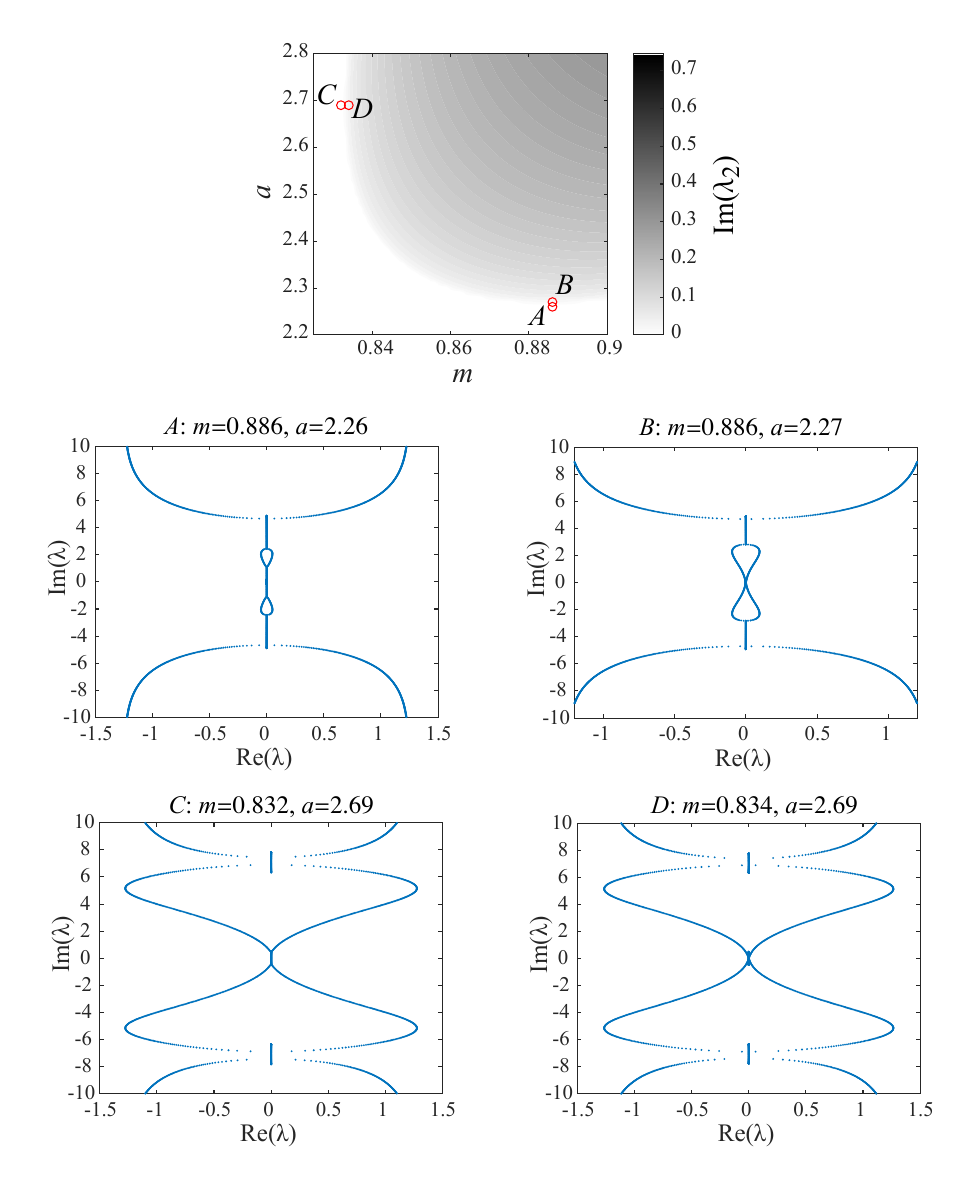}
\caption{Top panel: Fig.~\ref{fig:hyperbolicityIII}(a) zoomed in around the tip of the gray region where the Whitham system loses hyperbolicity in the case of quadratic nonlinearity ($\alpha=1$, $\beta=0$). Middle and bottom panels: spectra of the linear operator at the parameter values marked by points $A$, $B$, $C$ and $D$ in the top panel}
\label{fig:spectra_caseIII}
\end{figure}

We begin by examining the spectra in the case of quadratic
nonlinearity (Case III) with $(\alpha,\beta) = (1,0)$. The top panel
in Fig.~\ref{fig:spectra_caseIII} zooms in around the top of the gray
region in Fig.~\ref{fig:hyperbolicityIII}(a), where the Whitham
modulation system loses hyperbolicity. As we enter this region, we
expect to see the onset of modulational instability of periodic
traveling waves. To check this, we compute the linear spectrum for
periodic traveling waves at the parameter values $m$ and $a$
corresponding to points $A$ and $B$, located just outside and just
inside the gray region, respectively. The corresponding spectra are
shown in the two middle panels of
Fig.~\ref{fig:spectra_caseIII}. Eigenvalues away from the imaginary
axis correspond to instabilities. One can see that both solutions
exhibit short-wavelength instabilities ($|\lambda|>1$,
$\text{Re}(\lambda) \neq 0$), including superharmonic instabilities,
studied in detail in \cite{Bronski23}, which are associated with the
spectrum tending to infinity along vertical asymptotes. Note, however,
the spectrum at $A$ does not exhibit modulational instability (long
wavelength, $\lambda \sim 0$ with nonzero real part), while the
characteristic cross pattern of the spectrum near the origin
associated with modulational instability is clearly seen at $B$, the
point just inside the gray region. A similar emergence of modulational
instability can be seen as we transition from point $C$ just outside
the gray region to the point $D$ just inside it, as illustrated by the
bottom panels in Fig.~\ref{fig:spectra_caseIII}.

\begin{figure}
\centering
\includegraphics[width=\textwidth]{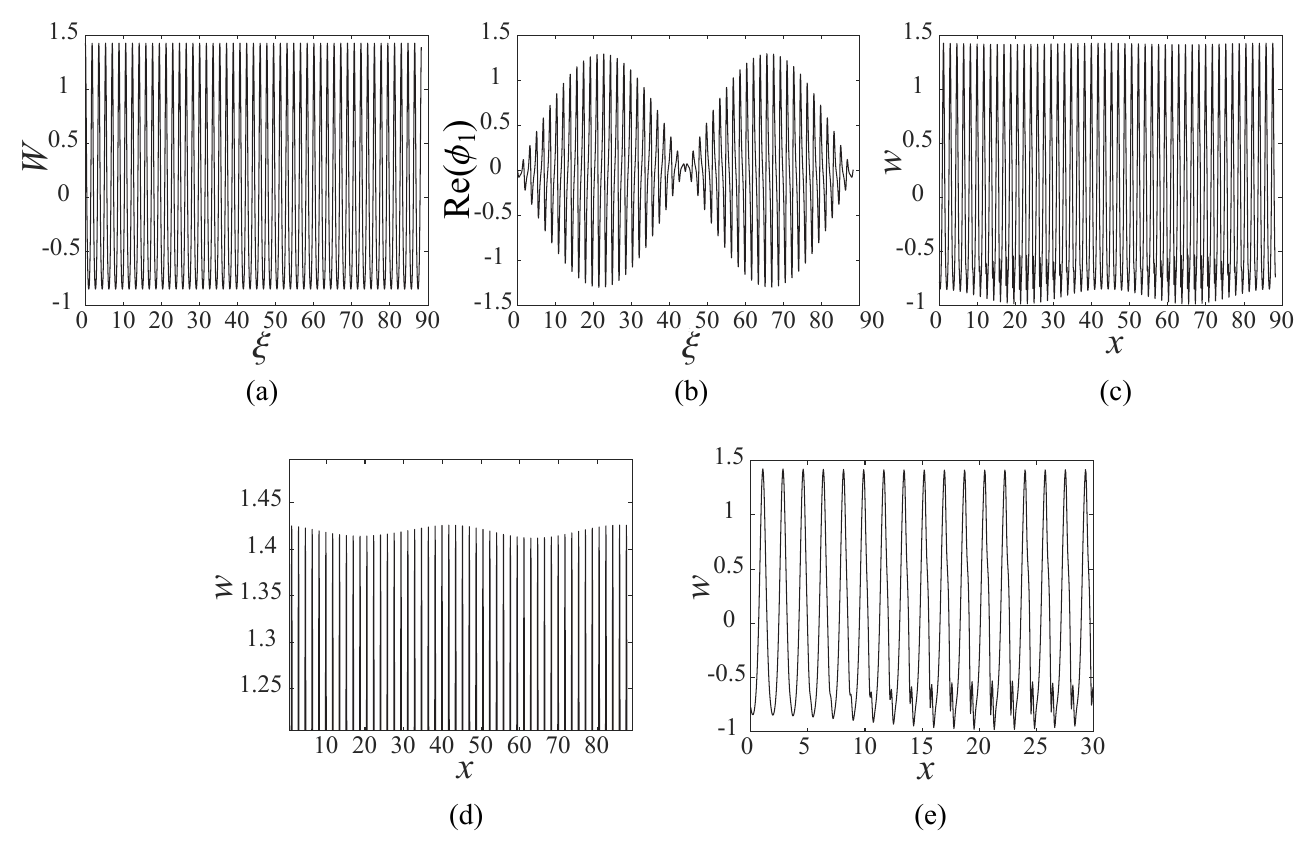}
\caption{(a) The strain component $W(\xi)$ of the
  periodic traveling wave solution at $m=0.886$, $a=2.27$. (b) Real
  part of the first component $\phi_1$ of the eigenfunction associated
  with the eigenvalue $\lambda=0.0019+0.0883i$. (c) Strain
  profile at $t=8$ in the dynamical simulation initiated by the
  traveling wave perturbed by the unstable mode, with perturbation
  amplitude $\eps=0.01$. (d) Panel (c) zoomed in around the top
  part. (e) Panel (c) zoomed in around the left part}
\label{fig:pert_dyn}
\end{figure}

To see the dynamical consequences of modulational instability, we
perturb the periodic traveling wave at $m=0.886$, $a=2.27$ (point $B$
in the top panel of Fig.~\ref{fig:spectra_caseIII}), the strain
component of which is shown in Fig.~\ref{fig:pert_dyn}(a), by adding
the unstable mode associated with the eigenvalue
$\lambda=0.0019+0.0883i$ (shown in Fig.~\ref{fig:pert_dyn}(b)),
with the amplitude of perturbation $\eps=0.01$. The strain snapshot at
$t=8$ is shown in panel (c) of Fig.~\ref{fig:pert_dyn}, with panels
(d) and (e) zooming in on the top and left side of the strain profile,
respectively. One can see that the strain profile clearly develops
modulational instability of the same wavelength as the perturbation
mode. Note, however, that in addition to this, a short-wavelength
instability develops as well, as can be seen in panel (e), in
agreement with the spectrum shown in the middle right panel of
Fig.~\ref{fig:spectra_caseIII} (point $B$). This instability
eventually leads to the blowup of the solution.

\begin{figure}
\centering
\includegraphics[width=0.75\textwidth]{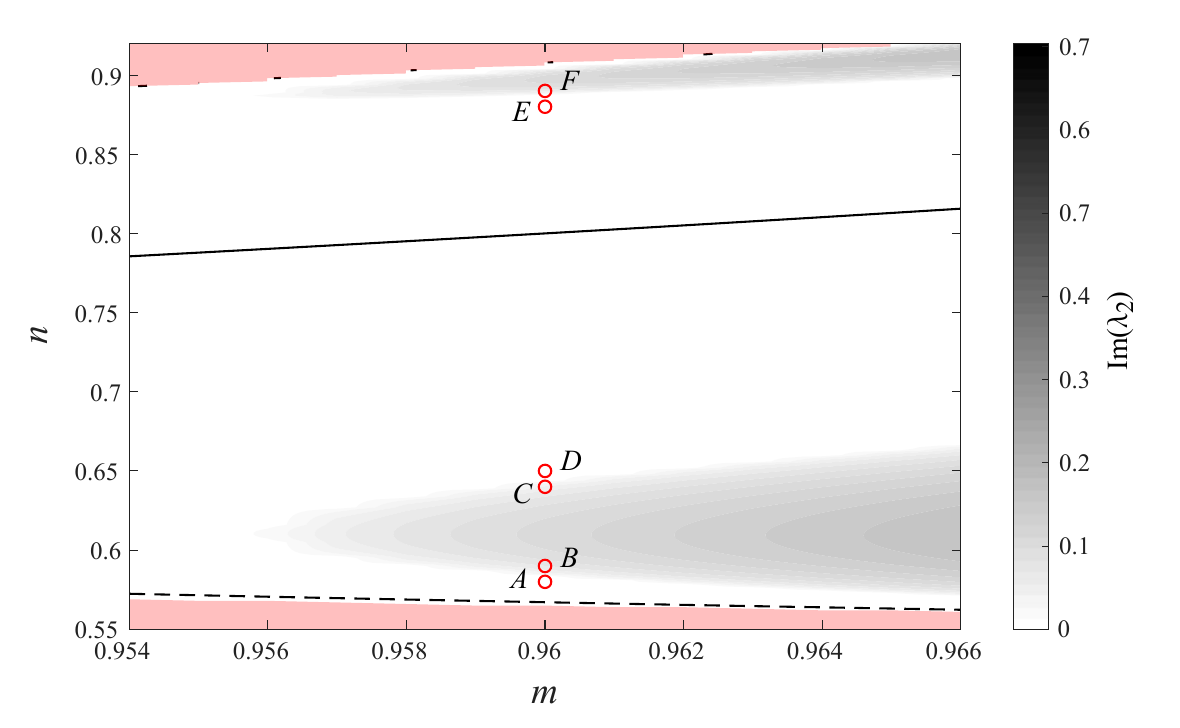}
\caption{Fig.~\ref{fig:hyperbolicityIIa}(d) ($a=1.3$)
  zoomed in near the tips of the gray regions where the Whitham system
  loses hyperbolicity in the case of cubic nonlinearity with
  $(\alpha,\beta) = (0,-1)$. Spectra shown in
  Fig.~\ref{fig:spectra_caseII} were computed at the parameter values
  marked by the points $A$-$F$}
\label{fig:hyperbolicityII_marked}
\end{figure}

\begin{figure}
\centering
\includegraphics[width=0.8\textwidth]{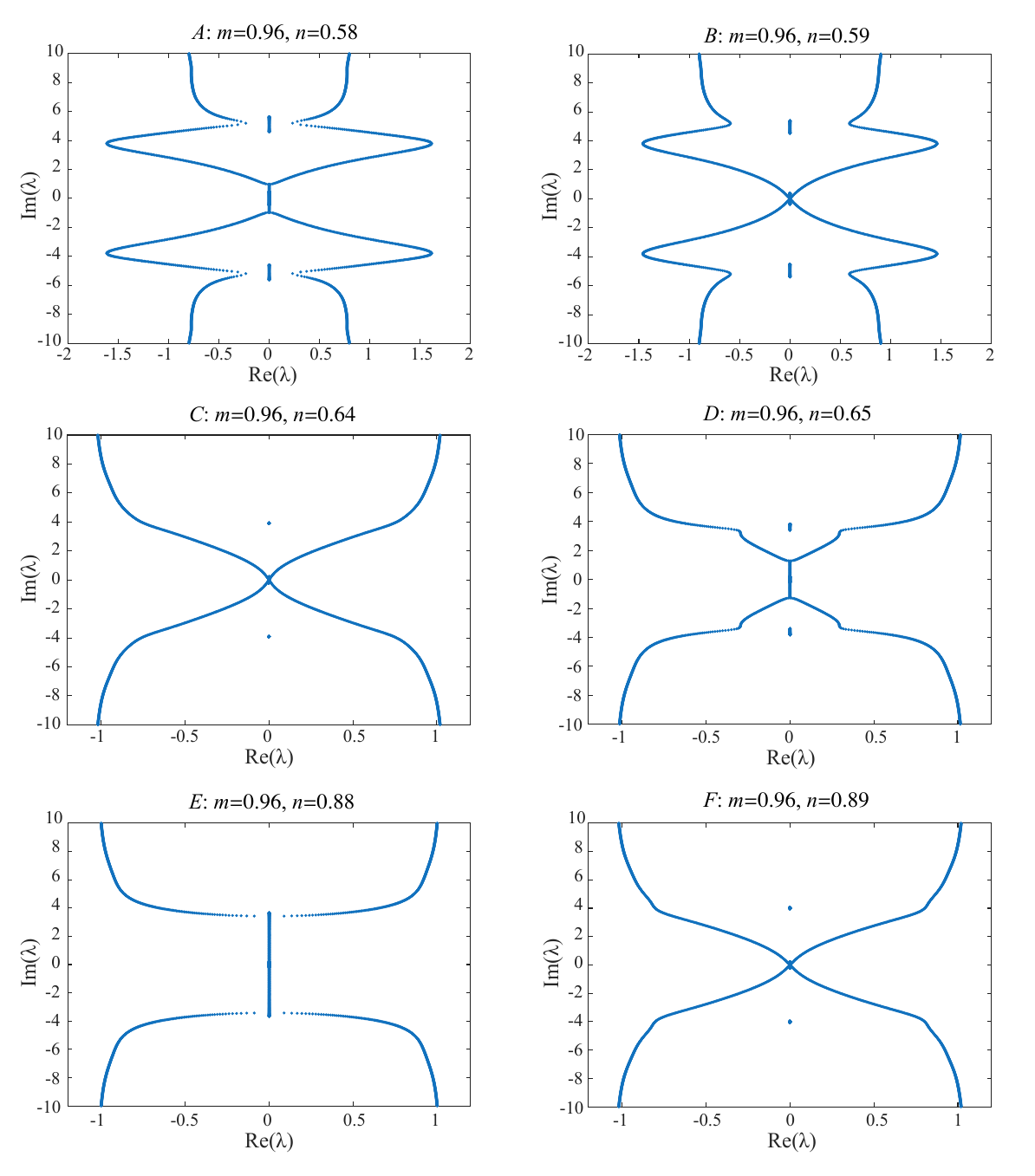}
\caption{Spectra of the linear operator at the parameter values marked by points $A$-$F$ in Fig.~\ref{fig:hyperbolicityII_marked}}
\label{fig:spectra_caseII}
\end{figure}

To check the onset of modulational instability in the case
$(\alpha,\beta) = (0,-1)$ (Case II), we consider the parameter values
at points $A$-$F$ in Fig.~\ref{fig:hyperbolicityII_marked}, which
zooms in on the area in Fig.~\ref{fig:hyperbolicityIIa}(d) (for
periodic traveling waves of amplitude $a=1.3$) near the tips of the
gray regions where the Whitham system loses hyperbolicity. The
corresponding spectra are shown in Fig.~\ref{fig:spectra_caseII}. One
can see that the modulational instability develops as we transition
from point $A$ outside the lower gray region to point $B$ inside it
(top panels). Point $C$ is still inside the region, and the
corresponding spectrum (left middle panel of
Fig.~\ref{fig:spectra_caseII}) shows modulational instability. This
instability disappears at point $D$ (right middle panel), which is
outside the gray region. The spectrum at point $E$, which is still in
the domain of strict hyperbolicity (see
Fig.~\ref{fig:hyperbolicityII_marked}), is shown in the left lower
panel of Fig.~\ref{fig:spectra_caseII}. One can see the modulational
instability developing again at point $E$ (right lower panel), which
is just inside the top gray region in
Fig.~\ref{fig:hyperbolicityII_marked}. Note that in all cases there
are short-wavelength instabilities.

\begin{figure}[h]
\centering
\includegraphics[width=\textwidth]{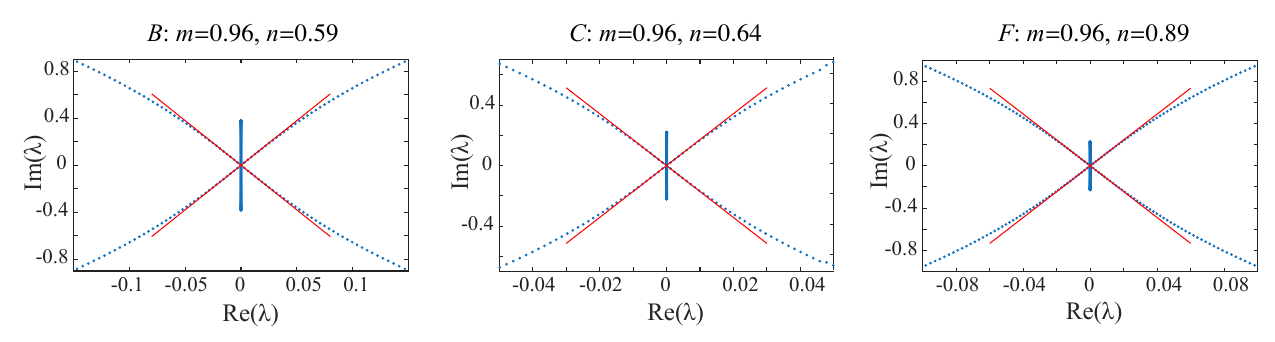}
\caption{Spectra of the linear operator at the parameter values marked by points $B$, $C$ and $F$ in Fig.~\ref{fig:hyperbolicityII_marked} zoomed in around the origin, shown together with red straight lines obtained from the analysis of the Whitham system. See the text for details}
\label{fig:spectra_caseII_MI}
\end{figure}

When the spectrum exhibits modulational instability, one can see a
characteristic cross structure near the origin. The slopes of this
structure can be estimated using the complex-valued Whitham
characteristic velocities $\sigma$ and $\bar{\sigma}$ that lead to the
instability. Indeed, for small wavenumber $k$ we have
$\text{Re}(\lambda) \approx k\text{Im}(\sigma)$ and
$\text{Im}(\lambda) \approx k(\text{Re}(\sigma)-c)$, so that the slope
$\text{Im}(\lambda)/\text{Re}(\lambda) \approx
(\text{Re}(\sigma)-c)/\text{Im}(\sigma)$, and the same is true for
$\bar{\sigma}$. To illustrate this, we show in
Fig.~\ref{fig:spectra_caseII_MI} the red lines with the corresponding
slopes superimposed with the spectra at the parameter values marked by
points $B$, $C$ and $F$ in Fig.~\ref{fig:hyperbolicityII_marked}
zoomed in around the origin.  
%

\begin{figure}[h]
\centering
\includegraphics[width=\textwidth]{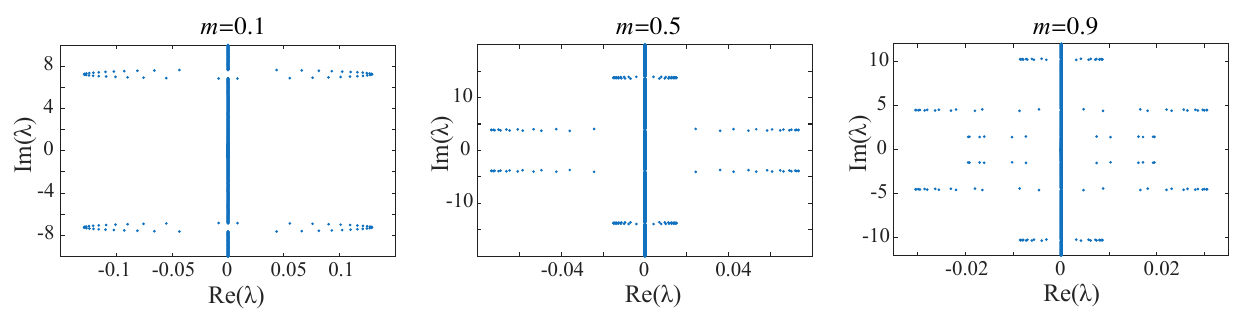}
\caption{Spectra of the linear operator at different values of parameter $m$ for Case I with $(\alpha,\beta)=(0,1)$ and real roots. Here $a=1$ and $n=0.5$}
\label{fig:spectra_caseI_real}
\end{figure}

Finally, we consider Case I with $(\alpha,\beta)=(0,1)$. Recall that
in the case when all roots are real, no loss of hyperbolicity is
observed, and thus all periodic traveling waves are expected to be
modulationally stable (cf.~Fig.~\ref{fig:hyperbolicityIa}). However,
other instability modes are still present, as illustrated in
Fig.~\ref{fig:spectra_caseI_real}, where we use the transformation
\eqref{eq:A_parameterization_transformation_caseIa} to the parameter
values \eqref{eq:A_parameterization_caseIandII}.
\begin{figure}
\centering
\includegraphics[width=0.8\textwidth]{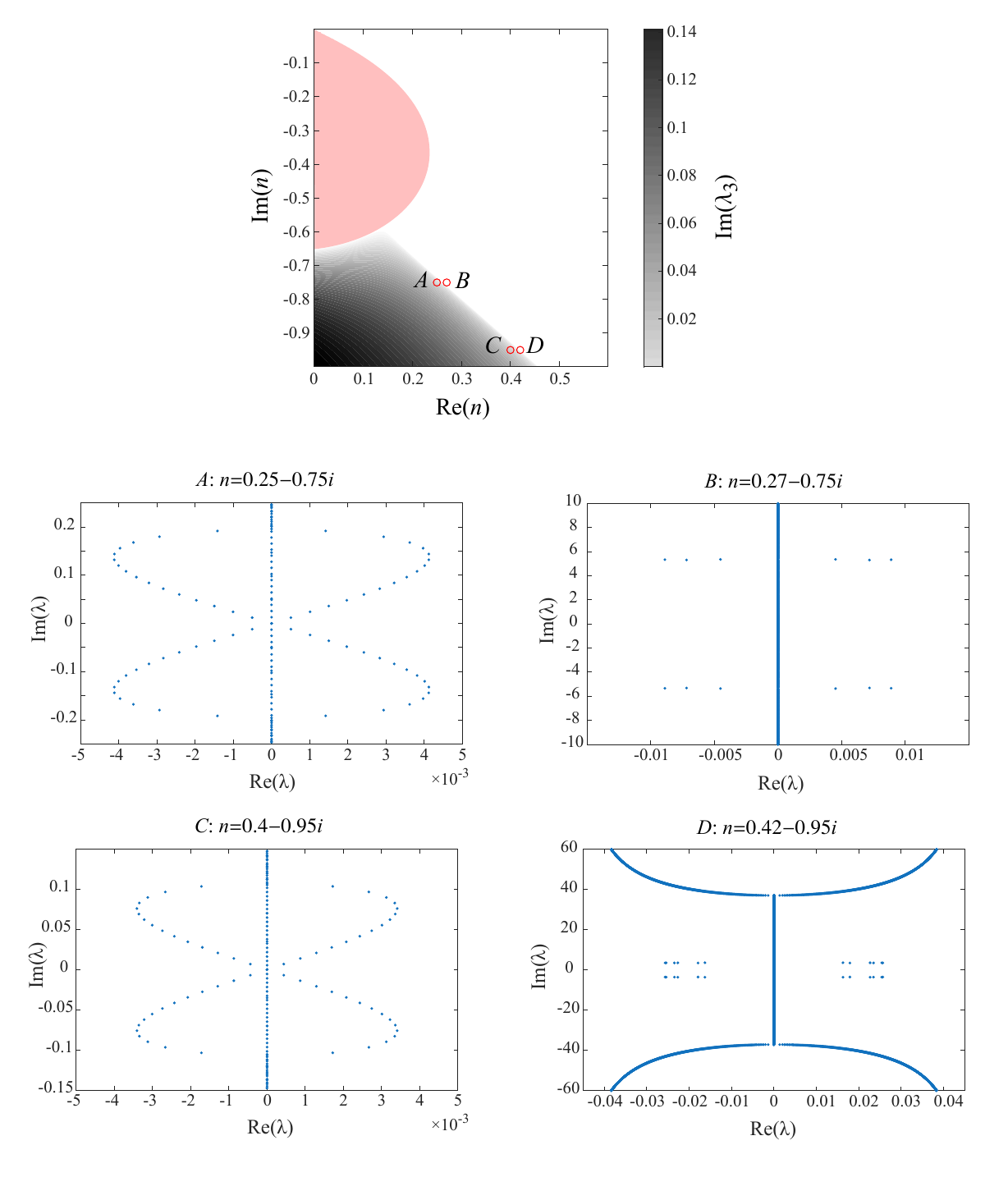}
\caption{Top panel: upper left part of $a=1$ panel in Fig.~\ref{fig:hyperbolicityIa_complex} for Case I with $(\alpha,\beta)=(0,1)$, real $w_{1,2}$ and complex $w_{3,4}$. Bottom panels: spectra of the linear operator at the parameter values marked by points $A$-$D$ in the top panel}
\label{fig:spectra_caseI_complex}
\end{figure}
In the case when $w_{3,4}$ are complex, we found that the Whitham
modulation equations are no longer hyperbolic in certain parameter
regimes as shown in Fig.~\ref{fig:hyperbolicityIa_complex}.  We use
the transformation \eqref{eq:transformation_caseI_complex} to present
the results in terms of \eqref{eq:A_param_caseI_complex} and recall
that the existence region for the periodic traveling waves is given by
\eqref{eq:tildeA_existence_caseIa}.  As shown in the top panel of
Fig.~\ref{fig:spectra_caseI_complex} (see also
Fig.~\ref{fig:hyperbolicityIa_complex} with $a = 1$), in this case
there is a gray region associated with the loss of hyperbolicity and
onset of modulational instability. This instability mode is evident in
the spectra at points $A$ and $C$ just inside this region (left middle
and bottom panels) and disappears in the spectra at points $B$ and $D$
just outside the gray region (right middle and bottom
panels). However, the spectra at $B$ and $D$ feature short-wavelength
instability.

Finally, we remark on stability of the periodic solutions
corresponding to the orbits shown in Fig.~\ref{fig:orbits1}(a-c) and
Fig.~\ref{fig:orbits2}.  All of these solutions are modulationally
stable but exhibit instabilities of shorter wavelength, with maximum
of $|\text{Re}\lambda|$ decreasing with amplitude for each set of
orbits.

\section{Concluding remarks}
\label{sec:conclusions}
In this work we derived the Whitham modulation equations for the
modified regularized Boussinesq equation that approximates the FPU
problem for cubic interaction force. We investigated the structure and
convexity of the Whitham system in the case of generally nonconvex
cubic nonlinearity, which yields explicit periodic traveling wave
solutions and enables complete analysis of the exact solitary-wave,
kink and harmonic limits. Away from these limits we conducted a
systematic study of the system's convexity based on numerical
computation of the eigenvalues and eigenvectors of the modulation
matrix. Our results clarify how the regions where the system is convex
depend on the amplitude, mean strain and other parameters of the
periodic traveling waves.

We confirmed the onset of modulational instability due to the loss of
hyperbolicity by computing the spectra of the linearized
operator and conducting numerical simulations initiated by
modulationally unstable waves perturbed along the corresponding
eigenmode. Although we did not perform an exhaustive study of linear
stability and mostly investigated solutions near the boundaries of the
convexity region, our results show that both modulationally stable and
unstable waves can exhibit short-wavelength instabilities that can
lead to solution blowup, in agreement with earlier findings
\cite{Pava13,Bronski23}. These instabilities are not captured by the
Whitham modulation system.

The results of this work pave the way for characterizing the jump
conditions and the oscillatory structure of dispersive shock waves
(DSWs) in the Boussinesq equation using the DSW fitting method
\cite{El05}, the analysis of exact solitary-wave limits \cite{Congy25}
and the direct computation of the integral curves of the modulation
system. The availability of the exact solitary-wave and harmonic-wave
limits also enables a more precise analysis of the interactions
between solitary waves or linear wavepackets and rarefaction waves or
DSWs \cite{Congy25}. Future work will also
include comparison of these results with simulations of FPU dynamics
for cubic nonlinearity. In particular, it will be interesting to
compare the regions of modulational and short-wavelength instabilities
in the discrete problem and its quasicontinuum approximation.\\

\noindent {\bf Acknowledgements.}
The authors would like to thank the
Isaac Newton Institute for Mathematical Sciences, Cambridge, for
support and hospitality during the programme ``Emergent phenomena in
nonlinear dispersive waves,'' where work on this paper was
initiated. This work was supported by EPSRC grant EP/V521929/1.  It
was also supported by the grants DMS-2306319 (M.A.H.) and DMS-2204880
(A.V.) from the U.S. National Science Foundation.


\bibliography{refs}

\end{document}